\documentclass[peerreview]{IEEEtran}
\IEEEoverridecommandlockouts

\usepackage[utf8]{inputenc}
\usepackage{stackengine}
\usepackage{amssymb}
\usepackage{amsmath}
\usepackage{mathrsfs}
\usepackage{ulem}
\usepackage{epsf,epsfig}
\usepackage{cite}
\usepackage{color,xcolor}
\usepackage{dsfont}
\usepackage{bbm}
\usepackage{amsthm}
\usepackage{physics}
\usepackage{tcolorbox}
\usepackage{wrapfig}
\usepackage{url}
\usepackage{graphicx}
\usepackage{eurosym}
\usepackage{enumitem}
\usepackage{nopageno}
\usepackage{fancyhdr}
\usepackage{mathtools}
\usepackage{stmaryrd}

\def\naturals{\mathbb{N}}
\def\reals{\mathbb{R}}

\def\fieldq{\mathcal{F}_{q}}
\def\Expectation{\mathbb{E}}

\definecolor{darkgreen}{rgb}{0,0.5,0}
\definecolor{darkmagenta}{rgb}{.5,0,.5}
\definecolor{darkblue}{rgb}{0,0,0.5}

\def\CalC{\mathcal{C}}
\def\CalD{\mathcal{D}}
\def\CalE{\mathcal{E}}

\def\CalH{\mathcal{H}}
\def\CalI{\mathcal{I}}
\def\CalJ{\mathcal{J}}

\def\CalL{\mathcal{L}}

\def\CalP{\mathcal{P}}

\def\CalT{\mathcal{T}}

\def\CalW{\mathcal{W}}
\def\CalX{\mathcal{X}}
\def\CalY{\mathcal{Y}}

\def\ulineA{\underline{A}}
\def\ulineB{\underline{B}}
\def\ulineC{\underline{C}}

\def\ulineK{\underline{K}}

\def\ulineM{\underline{M}}

\def\ulineR{\underline{R}}

\def\ulineW{\underline{W}}

\def\ulineb{\underline{b}}

\def\ulinek{\underline{k}}

\def\ulinem{\underline{m}}

\def\ulinew{\underline{w}}

\def\hatt{\hat{t}}

\def\haty{\hat{y}}

\def\CalC{\mathcal{C}}
\def\CalD{\mathcal{D}}
\def\CalE{\mathcal{E}}

\def\CalH{\mathcal{H}}
\def\CalI{\mathcal{I}}
\def\CalJ{\mathcal{J}}

\def\CalL{\mathcal{L}}

\def\CalP{\mathcal{P}}

\def\CalT{\mathcal{T}}

\def\CalW{\mathcal{W}}
\def\CalX{\mathcal{X}}
\def\CalY{\mathcal{Y}}

\def\hatY{\hat{Y}}

\def\ulineCalE{\underline{\mathcal{E}}}

\def\ulineCalI{\underline{\mathcal{I}}}

\def\ulineCalW{\underline{\mathcal{W}}}

\def\parsec{\par\noindent}
\def\med{\medskip\parsec}

\def\define{\mathrel{\ensurestackMath{\stackon[1pt]{=}{\scriptstyle\Delta}}}}

\def\olinee{\overline{e}}

\def\olinex{\overline{x}}

\def\olDelta{\overline{\Delta}}

\def\ScrA{\mathscr{A}}

\def\ScrE{\mathscr{E}}

\def\ScrM{\mathscr{M}}

\def\ScrT{\mathscr{T}}

\def\ttt{\mathtt{t}}

\def\ttT{\mathtt{T}}

\def\sfc{\mathsf{c}}

\def\sfB{\mathsf{B}}

\def\span{\mbox{span}}
\def\supp{\mbox{supp}}

\newcommand{\colBlue}[1]{\textcolor{blue}{#1}}

\usepackage{xparse}

\providecommand{\mathscr}[1]{\mathcal{#1}}
\providecommand{\mathds}[1]{\mathbb{#1}}


\newtheorem{fulldefinition}{Definition}
\newtheorem{fullproposition}{Proposition}
\newtheorem{fulltheorem}{Theorem}
\newtheorem{fulllemma}{Lemma}
\newtheorem{fullremark}{Remark}

\providecommand{\alphaorig}{\alpha_{\mbox{\tiny O}}}
\providecommand{\alphasim}{\alpha_{\mbox{\tiny S}}}
\providecommand{\alphaint}{\alpha_{\mbox{\tiny I}}}
\providecommand{\SimulationCPTP}{\boldsymbol{{\mathfrak{E}_{\mbox{\tiny sim}}}}}
\providecommand{\InterCPTP}{\boldsymbol{{\mathfrak{E}_{\mbox{\tiny int}}}}}
\providecommand{\boldphi}{\boldsymbol{\phi}}

\providecommand{\CalH}{\mathcal{H}}
\providecommand{\CalP}{\mathcal{P}}
\providecommand{\CalW}{\mathcal{W}}
\providecommand{\CalY}{\mathcal{Y}}

\providecommand{\ScrM}{\mathscr{M}}

\providecommand{\HlbSp}{\mathcal{H}}
\providecommand{\CalD}{\mathcal{D}}

\providecommand{\ScrE}{\mathscr{E}}

\providecommand{\CommK}{K}
\providecommand{\InfM}{M}
\providecommand{\Expectation}{\mathbb{E}}
\providecommand{\fieldq}{\mathbb{F}_{q}}
\providecommand{\define}{\mathrel{\stackrel{\Delta}{=}}}

\providecommand{\parsec}{\par\noindent}
\providecommand{\med}{\medskip\parsec}
\providecommand{\hatt}{\hat{t}}
\providecommand{\olDelta}{\overline{\Delta}}
\providecommand{\olinex}{\overline{x}}
\providecommand{\ttt}{\mathtt{t}}
\providecommand{\ttT}{\mathtt{T}}

\providecommand{\ulineb}{\underline{b}}
\providecommand{\ulb}{\uline{b}}
\providecommand{\ulineK}{\underline{K}}
\providecommand{\ulinek}{\underline{k}}
\providecommand{\ulinem}{\underline{m}}
\providecommand{\ulk}{\underline{k}}
\providecommand{\ulm}{\underline{m}}
\providecommand{\ulinebstarkm}{\ulineb^{*}_{\ulinek,\ulinem}}

\providecommand{\comment}[1]{}

\usepackage{microtype}
\makeatletter
\@ifpackageloaded{amsmath}{\allowdisplaybreaks[2]}{}
\makeatother

\def\olinescre{\overline{\mathscr{E}}}

\def\CommK{K}
\def\InfM{M}

\def\ulm{\underline{m}}
\def\SimulationCPTP{\boldsymbol{{\mathfrak{E}_{\mbox{\tiny sim}}}}}
\def\InterCPTP{\boldsymbol{{\mathfrak{E}_{\mbox{\tiny int}}}}}
\def\boldphi{\boldsymbol{\phi}}

\begin{document}

\title{\huge{Distributed Instrument Simulation with Quantum Side Information in the One-Shot Regime}}

\author{%
  \IEEEauthorblockN{Igor BERNARD and Arun PADAKANDLA}
}

\maketitle

\begin{abstract}
Three distributed parties, two transmitters (Txs) and a receiver (Rx), hold one component each of a tripartite quantum state \(\rho^{A_1A_2C}\). The goal is to simulate the action of a separable instrument acting on the \(A_1\) and \(A_2\) components, with the Rx recovering the classical outcome. To enable this, each Tx \(k\) can transfer bits on a noiseless bit pipe and share randomness at rates \(R_k\) and \(C_k\), respectively, with the Rx. Undertaking a Shannon-theoretic study, we characterize two new sets of inner bounds. The first set, derived for the one-shot regime, is based on instrument simulation protocols built using unstructured IID codes, while the second set, derived for the asymptotic regime, relies on coset codes and new decoding POVMs. The first set of bounds recovers current known inner bounds for instrument and measurement simulation in all previously studied scenarios. Our protocols are based on likelihood POVMs, and our analysis leverages Sen's smooth multiparty covering and simultaneous decoding, while handling the distributed-component scenario via a compatible operator sliding trick.
\end{abstract}
\section{Introduction}
\label{Sec:Introduction}
Building on prior works \cite{197109IJP_Gro,200005PhyA_MasPop,200101PhyA_WinMas}, Winter elegantly formulated \cite{200401CommMathPhy_Win} the important question of quantifying the information content in the outcome of a measurement into what we now refer to as the measurement compression problem (MCP). At its core is quantum classical covering (QC) - a tool central to several problems \cite{200501TIT_Dev,201301TIT_DatHsiWil,201405TIT_BenDevHarShoWin}. These connections and broader applicability of QC covering in several problems have inspired MCP studies \cite{2012MMJPhyA_WilHayBusHsi,202202TIT_AtiHeiPra,202206ISIT_ChaPadSen,202504TIT_AtiPra,201909TIT_AnsJaiWar} in diverse network scenarios. Here, we address a network scenario involving general measurements and quantum side information (QSI) in the one-shot regime.

\begin{figure}[t]
\centering
\includegraphics[width=0.70\textwidth]{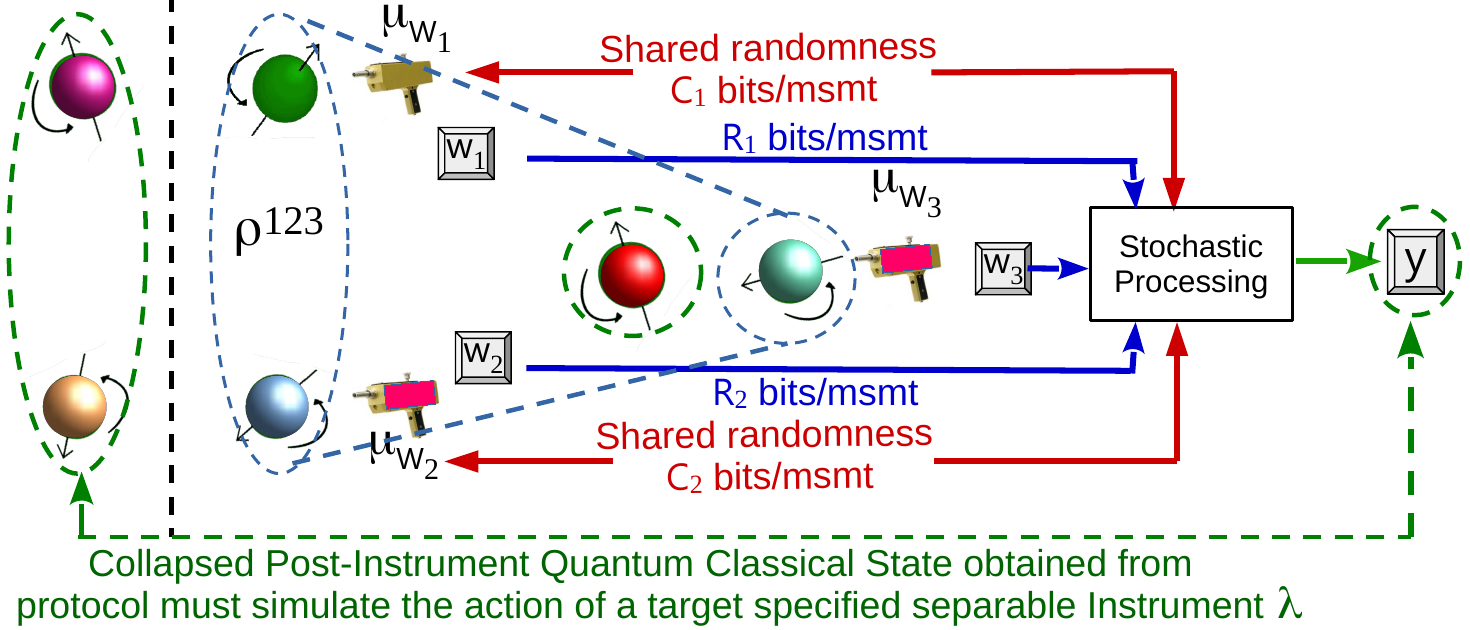}
\caption{Illustration of the distributed instrument simulation with QSI setting.}
\label{Fig:DistsInstQSI}
\end{figure}

Three distributed parties -- two transmitters (Txs), Alice$_1$ and
Alice$_2$, and an Rx, Charlie -- hold one component each of a possibly entangled tripartite quantum state $\rho^{A_1A_2C}$ (Fig.~\ref{Fig:DistsInstQSI}). The goal is to simulate the action of a \textit{separable} \textit{instrument} acting on $A_{1},A_{2}-$components with Charlie recovering the classical outcome. To accomplish this, each Tx $k$ can transfer bits on a noiseless bit pipe of rate $R_{k}$ to the Rx and share common random bits at rate $C_{k}$ with the latter. We undertake a Shannon theoretic study in the one-shot regime and characterize a new inner bound (Theorem~\ref{Thm:DistInstSimltnOneShot}) on the rate quadruples $(R_{k},C_{k}: k \in [2])$ for which this is possible. Next, leveraging the ensemble of \textit{coset codes} which are known to yield larger inner bounds in distributed component scenarios, we present a new inner bound (Theorem~\ref{Thm:AsymptoticCosetInstSimulation}) in the asymptotic regime that utilizes Rx QSI. 

To place our formulation and findings in perspective, let us discuss prior work. In their expository article, Wilde et.~al.~\cite{2012MMJPhyA_WilHayBusHsi} modeled a Rx's knowledge of quantum side information (QSI) and proved it requires fewer bits to simulate the classical outcome. Recognizing MCP's importance, Anshu.~et.~al \cite{201909TIT_AnsJaiWar} spearheaded its study in the one-shot regime. They employed convex-split \cite{201709PRL_AnsDevVamJan} - a key tool that has spurred much research \cite{202511TIT_CheGao}. While \cite{201909TIT_AnsJaiWar} modeled Rx QSI, the inherent difficulty of one-shot analysis had restricted \cite{201909TIT_AnsJaiWar}'s formulation to requiring resources catalytically, which was overcome in \cite{202206ISIT_ChaPadSen}.

The next important step involved simulating separable measurements on components of a bipartite state distributed amidst two transmitters (Tx) studied by Atif et.~al.~\cite{202202TIT_AtiHeiPra}. The analysis being sufficiently challenging, \cite{202202TIT_AtiHeiPra} resort to Winter's typicality based simulation protocol lending their results inapplicable in the one-shot regime as is evident from their restricted one-shot study and sub-optimal bounds \cite[Theorem~3]{202202TIT_AtiHeiPra}. The design of two separate simulation measurement for the distributed component scenario introduces new challenges particularly in the one-shot regime owing to lack of simplifying techniques such as time sharing. \cite{2026MMTIT_Pad} develops an alternate protocol based on likelihood POVMs and tackles the one-shot distributed MCP to derive inner bound analogous to the asymptotic case.

Two observations are evident from these pursuits \cite{2012MMJPhyA_WilHayBusHsi,202202TIT_AtiHeiPra,202206ISIT_ChaPadSen,202504TIT_AtiPra,201909TIT_AnsJaiWar}. Firstly, MCP is a central problem in quantum info.~th.~whose pursuit has yielded tools applicable more broadly. Secondly, network scenarios involving additional components and/or QSI lend the analysis challenging thus spurring newer approaches, particularly in the one-shot regime due to its generality. These motivate our formulation. Moreover, as we discuss, we relax a common underlying restriction in prior works \cite{2012MMJPhyA_WilHayBusHsi,202202TIT_AtiHeiPra,201909TIT_AnsJaiWar}.
The focus of \cite{200401CommMathPhy_Win,2012MMJPhyA_WilHayBusHsi,202202TIT_AtiHeiPra,202504TIT_AtiPra,201909TIT_AnsJaiWar} has been to quantify \textit{correlations} established between the post-measurement state and the classical outcome by a measurement's action. This has led to two assumptions in their formulations. Firstly, the class of measurements studied is restricted to positive operator valued measurements (POVM). Secondly, since the latter leaves the accessible component's phase unspecified, and moreover, correlations can be captured via the inaccessible reference, most studies ignore the accessible component. These clever assumptions, while not altering the final rate characterizations, have simplified the necessary mathematical tools and analysis. Indeed, a slew of inequalities and techniques applicable only for self-adjoint/positive operators fall within the purview of the analysis. However, not requiring preservation of the accessible component makes these solutions insufficient for tasks such as purity \cite{200707PRA_KroDev} and entanglement distillation \cite{199601PRL_BenBraPopSchSmoWoo, 200501MPES_DevWin} since the latter requires phase preservation of the accessible component. Moreover, their solutions are closely linked to that of MCP.

An \textit{instrument} \cite[Sec.~4.6.8]{BkWilde_2017} is a general measurement that models a post-measurement quantum \textit{and} classical state not merely the latter as in a POVM. This motivates our first distinguishing aspect - formulation involving distributed separable instruments. Separability ensures distributed simulation via classical post processing (Definition~\ref{Defn:SeparableInstrument}).
Next, we model Rx QSI. True to our aforementioned second observation, this adds a new complexity. To efficiently utilize Rx QSI to compress rates, we must perform simultaneous decoding of the bin indices as in a CQ multiple access channel - an aspect unaddressed in \cite{200401CommMathPhy_Win,2012MMJPhyA_WilHayBusHsi,202202TIT_AtiHeiPra,202504TIT_AtiPra,201909TIT_AnsJaiWar}. We go furthermore and consider design of instruments based on coset codes - a development analogous to \cite[Sec.~V]{202407ISIT_PadWar}, \cite{202504TIT_AtiPra}. Essentially, the early work of K\"orner and Marton \cite{197903TIT_KorMar}, followed by \cite{200912TIT_KriPra,202301TIT_AtiPadPra} have proven that one can recover compressive bivariate functions of distributed sources more efficiently by designing coset codes endowed with joint algebraic properties. When the target instrument simulation admits an efficient decomposition via post processing of a bivariate function (See the stochastic matrix in the triples contained in $\ScrT_{\oplus}(\CalJ,\rho^{A_1A_2C})$ in Sec.~\ref{SubSec:InstSimltnAsymptoticCoset}), then instrument simulation protocols via coset codes yield high rates since the binning rate can be increased.

{We conclude with remarks on tools and novelty. The tools at the core of our analysis are Sen's fully smooth multi-party covering and Sen's tilting smooth and augmentation based simultaneous decoding that we leverage for the decoder analysis. In addition to these, the identification of the right proxy states and crucial order of sliding compatible operators couples with the right unitaries to transform the references - referred to as compatible operator sliding facilitates analysis of the distributed instrument scenario and overcome challenges encountered in \cite{202202TIT_AtiHeiPra}.

\section{Preliminaries and Problem Statements}
\label{Sec:PrelimsProblemStatement}
\subsection{\textbf{Notations}}
We supplement notation in \cite{BkWilde_2017} with the following. For $n \in \naturals$, we let $[n] \define \{1,\cdots,n\}, [\overline{n}]\define\{0\}\cup [n]$.
$\CalH_{A}$ and $\CalH_{X}$ will denote state spaces of the accessible and inaccessible (reference) components of our quantum systems respectively. Throughout, all Hilbert spaces are assumed to be finite-dimensional and sets to be of finite cardinality. An \underline{underline} denotes an appropriate aggregation of related objects. For ex.~if $\CalH_{A_{1}},\CalH_{A_{2}}$ are Hilbert spaces, $\CalH_{\ulineA}$ denotes $\CalH_{A_{1}}\otimes\CalH_{A_{2}}$, whereas when $k_{j} \in [K_{j}]$ for $j \in [2]$ are elements in index sets, $\ulinek =(k_{1},k_{2})$ or we let $p_{Y|\ulineW}$ denote $p_{Y|W_{1}W_{2}}$. $\mathcal{L}(\mathcal{H})$, $\mathcal{P}(\mathcal{H})$ and $\mathcal{D}(\mathcal{H})$ denote linear, positive and density operators acting on $\mathcal{H}$. For any $\ket{x} \in \CalH$, we let $\ket{\olinex}\in \CalH$ denote complex conjugation with respect to some fixed orthonormal basis of $\CalH$. See \cite[App.~A]{200401CommMathPhy_Win}. For state $\rho \in \CalD(\CalH)$ with spectral decomposition $\rho=\sum_{x} \alpha_{x}\ketbra{e_{x}}$, $\ket{\varphi_{\rho}} =  \sum_{x} \sqrt{\alpha_{x}}\ket{\olinee_{x}}\otimes\ket{e_{x}} \in \CalH \otimes \CalH$ denotes the \textit{canonical purification} and we let $\varphi_{\rho}\define \ketbra{\varphi_{\rho}}$. For a stochastic matrix $(p_{Y|W}(y|w):(w,y)\in \mathcal{W}\times \mathcal{Y})$, we let $\olinescre^{{Y|W}}_{p}(\cdot),\ScrE^{{Y|W}}_{p}(\cdot)$ denote the CPTP map $\olinescre^{Y|W}_{p}(a)
\define
\sum_{(w,y)\in\CalW\times\CalY}
p_{Y|W}(y|w)\,
\ket{w}\!\mel{w}{a}{w}\!\bra{w}
\otimes \ketbra{y},$ and set $\ScrE^{Y|W}_{p}\define \Tr_{W}\circ \olinescre^{Y|W}_{p}.$ CP, TNI, TP, QSI, CQMAC abbreviate completely positive, trace non-increasing, trace preserving, quantum side information, classical–quantum multiple-access channel respectively.
\colBlue{}
\subsection{\textbf{Instrument Simulation : Objects and Problem Statements}}
\label{SubSec:InstSimltnProbStatements}
\begin{fulldefinition}[Quantum instrument]
\label{Defn:QuantumInstrument}
A quantum instrument with input system \(A\), output system \(B\), and
finite outcome alphabet \(\CalY\), is a collection $\CalJ=\{\CalJ_y:\CalL(\CalH_A)\to\CalL(\CalH_B)\}_{y\in\CalY}$ of CP TNI maps such that \(\sum_{y\in\CalY}\CalJ_y\) is TP. Its action is identified with the CPTP map $\CalJ(\rho)\define \sum_{y\in\CalY}\CalJ_y(\rho)\otimes\ketbra{y}_Y .$

Equivalently, there exist a finite-dimensional environment \(\CalH_E\)
and operators \(M_y:\CalH_A\to\CalH_B\otimes\CalH_E\), \(y\in\CalY\),
such that
\begin{eqnarray}
\label{Eqn:QuantInstrument}
\CalJ(\rho)
=
\operatorname{Tr}_E\!\left[
\sum_{y\in\CalY}
M_y\rho M_y^\dagger\otimes\ketbra{y}_Y
\right],
\end{eqnarray}

with $\sum_{y\in\CalY}M_y^\dagger M_y=I_A.$ Moreover, by enlarging \(\CalH_E\) if necessary, one may assume
\(\dim\CalH_A\leq \dim(\CalH_B\otimes\CalH_E)\).
\end{fulldefinition}
\begin{fullproposition}[Operator-sum realization]
\label{Prop:InstrumentOperatorSum}
Every instrument in Definition~\ref{Defn:QuantumInstrument} admits
the representation \eqref{Eqn:QuantInstrument}.
\end{fullproposition}
\begin{proof}
For each \(y\in\CalY\), choose a Kraus representation $\CalJ_y(\rho)=\sum_{x=1}^{r_y}M_{x,y}\rho M_{x,y}^\dagger .$ Let \(r=\max_{y\in\CalY}r_y\), pad the Kraus families with zero operators,
and let \(\CalH_E\) have orthonormal basis
\(\{\ket{x}_E:x\in[r]\}\). Define $M_y\ket{\alpha}
=
\sum_{x=1}^r
\bigl(M_{x,y}\ket{\alpha}\bigr)\otimes\ket{x}_E, \ket{\alpha}\in\CalH_A .$ Then $M_y\rho M_y^\dagger
=
\sum_{x,x'=1}^r
M_{x,y}\rho M_{x',y}^\dagger\otimes\ketbra{x}{x'}_E,$
and hence $\operatorname{Tr}_E[M_y\rho M_y^\dagger]
=
\sum_{x=1}^r M_{x,y}\rho M_{x,y}^\dagger
=
\CalJ_y(\rho).$ Therefore $\sum_{y\in\CalY}\CalJ_y(\rho)\otimes\ketbra{y}_Y
=
\operatorname{Tr}_E\!\left[
\sum_{y\in\CalY}
M_y\rho M_y^\dagger\otimes\ketbra{y}_Y
\right].$ Finally, $\sum_{y\in\CalY}M_y^\dagger M_y
=
\sum_{y\in\CalY}\sum_{x=1}^r M_{x,y}^\dagger M_{x,y}
=
I_A,$
because \(\sum_{y\in\CalY}\CalJ_y\) is trace preserving. Enlarging
\(\CalH_E\), if needed, gives
\(\dim\CalH_A\leq\dim(\CalH_B\otimes\CalH_E)\).
\end{proof}
Our focus will be on simulating a separable instrument defined below.
\begin{fulldefinition}
 \label{Defn:SeparableInstrument}
 An instrument $\mathcal{J}:\ \CalL(\CalH_{\ulineA} ) \longrightarrow \CalL(\CalH_{\ulineB} \otimes \mathbb{C}^{\mathcal{Y}})$ acting\footnote{Recall our \underline{underline} notation. $\CalH_{\ulineA}$ denotes $\CalH_{A_{1}}\otimes\CalH_{A_{2}}$ and so on.} on a bipartite state is separable if there exists, for $i \in [2]$ instruments $\CalI^{(i)}: \CalL(\CalH_{A_{i}}) \longrightarrow \CalL(\CalH_{B_i} \otimes \mathbb{C}^{\mathcal{W}_{i}}) $ and a stochastic matrix $p_{Y|W_{1}W_{2}}$ such that, for each $y \in \CalY$ we have
 \begin{eqnarray}
 \label{Eqn:SeparableInstrument}
\CalJ_{y} = \sum_{w_{1},w_{2}}p_{Y|W_{1}W_{2}}(y|w_{1},w_{2}) \CalI_{w_{1}}^{(1)} \otimes \CalI_{w_{2}}^{(2)}.
 \end{eqnarray}
\end{fulldefinition}

We now describe the distributed setting, illustrated in
Fig.~\ref{Fig:DistsInstQSI}. Alice$_i$ holds system $A_i$ of the
tripartite state $\rho^{A_1A_2C}$, while Charlie holds the QSI system
$C$. The goal is to simulate the action of the separable instrument
\[
\mathcal{J}:\ \CalL(\CalH_{A_{1}}\otimes \CalH_{A_{2}} )
\longrightarrow
\CalL(\CalH_{B_1}\otimes \CalH_{B_2}\otimes\mathbb{C}^{\mathcal{Y}}),
\]
so that Charlie obtains the classical outcome $Y$, while the
post-instrument quantum systems remain with Alice$_1$ and Alice$_2$.
Each Alice$_i$ communicates to Charlie through a limited-rate classical
link and shares statistically independent common randomness with him.

Since $\CalJ$ is separable, it suffices for Alice$_i$ to simulate the
local instrument $\CalI^{(i)}$, after which Charlie applies the
stochastic post-processing $p_{Y|W_1W_2}$. Instead of implementing the
full instruments with outcome alphabets $\CalW_1$ and $\CalW_2$, each
Alice$_i$ applies an encoding instrument with only $M_i\ll|\CalW_i|$
outcomes and sends the resulting message to Charlie. Using the two
messages, the shared randomness and his QSI $C$, Charlie reconstructs
outcomes $\hat W_1,\hat W_2$ and produces $\hatY$ according to
$p_{Y|W_1W_2}$. The induced state on the post-instrument systems,
Charlie's QSI, the classical outcome, and the inaccessible reference
$X$ must be indistinguishable from the target state
\begin{eqnarray}
\label{Eqn:TargetState}
\CalT \define
\left(\mathrm{id}_{XC}\otimes \CalJ\right)
\{\ketbra{\varphi^{XA_{1}A_{2}C}}\}.
\end{eqnarray}
We now formalize the protocol and the corresponding simulation criterion.

\begin{fulldefinition}
\label{Defn:DistributedPOVMProtocol}
{A $(\ulineK,\ulineM,\ulineCalE,\CalD)$ one-shot instrument simulation (IS) protocol consists of (i) bank of $K_{i}$ instruments $\mathcal{E}^{(i)}_{k_i}:\ \CalL(\CalH_{A_i})\longrightarrow \CalL(\CalH_{B_i}\otimes \mathbb{C}^{M_i}) : k_{i} \in [K_{i}]$ each with $M_{i}$ outcomes for $i \in [2]$, and a (ii) decoder instrument $\CalD_{\ulinek} : \CalL(\CalH_{C}\otimes\mathbb{C}^{M_1}\otimes \mathbb{C}^{M_2})
\longrightarrow \CalL(\CalH_{C}\otimes\mathbb{C}^{|\mathcal{Y}|})$. \textbf{$\CalJ $'s action on $\rho^{A_{1}A_{2}C}$ can be $\eta$-simulated with (communication) cost $(\ulineR,\ulineC)$} if there exists a $(\ulineK,\ulineM,\ulineCalE,\CalD)$ one-shot IS protocol for which $\log K_{i} \leq C_{i}$, $\log M_{i} \leq R_{i}$ for $i\in [2]$ and state $\CalT$ in \eqref{Eqn:TargetState} satisfies}
\begin{eqnarray}
 \label{Eqn:OneShotSimulationTraceConstraint}
 \norm{\!\left(\mathrm{id}_{XC}\otimes \!\!\!\sum_{k_{1},k_{2}}\!\frac{\left[\CalD_{\ulinek}\! \circ\! \{ \ScrE^{(1)}_{k_{1}} \!\otimes \ScrE^{(2)}_{k_{2}} \} \right]}{K_{1}K_{2}} \!\right)\!\!(\varphi^{XA_{1}A_{2}C}) -\CalT }_{1} \!\leq \eta.\!\!\!\!
 \nonumber
 \end{eqnarray}
\textbf{$\CalJ$'s action on $\rho^{A_{1}A_{2}C}$ can be perfectly simulated with (communication) cost $(\ulineR,\ulineC)$} if for every $\eta>0$, there exists $N_{\eta}$ such that for all $n \geq N_{\eta}$, $\otimes_{t=1}^{n}{\CalJ}$'s action on $\otimes_{t=1}^{n}\rho^{A_{1}A_{2}C}$ can be $\eta$-simulated with cost $(n\ulineR,n\ulineC)$.
\end{fulldefinition}
\section{Multi-Party Quantum Instrument Simulation}
\label{Sec:InstSimulationWithIIDCodes}
In this section, we present two inner bounds (Theorems~\ref{Thm:DistInstSimltnOneShot}, \ref{Thm:AsymptoticCosetInstSimulation}) for the problem of instrument simulation. We exploit the common structural characterization that these two inner bounds possess to provide, thereby compromising readability slightly, a unified characterization. We therefore begin with preliminaries before addressing one-shot simulation with QSI via IID codes in Sec.~\ref{SubSec:OneShotISWithIIDCodes}. Our next Definition~\ref{Defn:TestChannels} enables us characterize all possible ``test channels'' or single-letter instruments that we can utilize in designing our instrument simulation protocol.
\begin{fulldefinition}
 \label{Defn:TestChannels}
 For instrument $\mathcal{J}:\ \CalL(\CalH_{\ulineA} ) \longrightarrow \CalL(\CalH_{\ulineB} \otimes \mathbb{C}^{\mathcal{Y}})$ and state $\rho^{A_{1}A_{2}}$, let $\ScrT(\CalJ,\rho^{A_{1}A_{2}})$ denote the collection $(\ulineCalW,\ulineCalI,p_{YU|\ulineW})$ of all triples, where (i) $\CalI^{(i)}: \CalL(\CalH_{A_{i}}) \longrightarrow \CalL(\CalH_{B_i} \otimes \mathbb{C}^{\mathcal{W}_{i}}) $ are instruments for $i \in [2]$, a stochastic matrix $p_{Y|W_{1}W_{2}}$ such that \eqref{Eqn:SeparableInstrument} holds for all $y \in \CalY$ and a decomposition $p(w_1,w_2)=\operatorname{Tr}\left[(\CalI_{w_{1}}^{(1)}\otimes\CalI_{w_{2}}^{(2)})\{\rho^{A_{1}A_{2}}\}\right]=\sum_{u}{p(w_1|u)p(w_2|u)p(u)}$ (together inducing a joint distribution $P_{W_1W_2YU}$). For a triple $(\ulineCalW,\ulineCalI,p_{YU|\ulineW})$ and a state $\rho^{A_{1}A_{2}C}$, we denote $\ulineCalI = \left[\CalI^{(1)}\otimes \CalI^{(2)}\right]$ and let
\begin{eqnarray}
\label{Eqn:TestChannelJointStates}
 \sigma^{XB_{1}B_{2}CW_{1}W_{2}Y_{1}Y_{2}U} \!\define\! \left(\!\mathrm{id}_{XC}\otimes\olinescre^{YU|{\ulineW}}_{p}\!\otimes\! \ulineCalI\right)\!\!\{{\varphi^{XA_{1}A_{2}C}}\}
 \\
 \label{Eqn:TestChannelIndividualStates}
  \sigma_{i}^{XB_{i}A_{3-i}CW_{i}} \!\define\! \left(\!\mathrm{id}_{XA_{3-i}C}\otimes \CalI^{(i)}\right)\!\!\{{\varphi^{XA_{1}A_{2}C}}\}.
\end{eqnarray}
\end{fulldefinition}

\subsection{Instrument Simulation with IID Codes : One-Shot Regime}
\label{SubSec:OneShotISWithIIDCodes}
Our first finding is an inner bound to distributed instrument simulation with QSI in the one-shot regime. Towards characterizing the same, we begin by identifying the one-shot information quantities appearing in the pre FME bounds.

\begin{fulldefinition}
\label{Defn:OneShotPreFMInfoQuantities}
For state $\rho^{A_1A_2C}$,
$(\ulineCalW,\ulineCalI,p_{YU|\ulineW}) \in
\ScrT(\CalJ,\rho^{A_{1}A_{2}})$
(Definition~\ref{Defn:TestChannels}) and $\epsilon>0$, set
$\epsilon_{5}= \log \frac{1}{\epsilon}$ and
$\underline{\sfB^{\epsilon}}
\define(\sfB^{\epsilon}_{1},\ldots,\sfB^{\epsilon}_{8})$, where
\begin{eqnarray}
 \begin{array}{l}
\sfB^{\epsilon}_{1}  = I_{2}^{\varepsilon}(W_1\!:\!CA_{2}X)_{\sigma_1}\!\!+2\epsilon_{5},~
\sfB^{\epsilon}_{2}  = I_{2}^{\varepsilon}(W_2\!:\!CA_{1}X)_{\sigma_2}\!\!+2\epsilon_{5}\!\!\\
\sfB^{\epsilon}_{3} = \displaystyle I_{2}^{\varepsilon}(W_1:YC\ulineB X)_{\sigma}+2\epsilon_{5} ,~\\
\sfB^{\epsilon}_{4} = I_{2}^{\varepsilon}(W_2:YC\ulineB X)_{\sigma}+2\epsilon_{5}\!\!
\\
\sfB^{\epsilon}_{5} = D_{2}^{\varepsilon}(\sigma^{XC\ulineB\ulineW Y}||\sigma^{W_{1}}\otimes \sigma^{W_{2}}\otimes\sigma^{XC\ulineB Y})_{\sigma}\!\!+2\epsilon_{5},~\\
\sfB^{\epsilon}_{6} = I_H^\varepsilon(W_1:W_2C\,|\,U)_\sigma \;-\; 2 -\epsilon_{5},~\\
\sfB^{\epsilon}_{7} = I_H^\varepsilon(W_2:W_1C\,|\,U)_\sigma \;-\; 2 -\epsilon_{5},~\\
\sfB^{\epsilon}_{8} = I_H^\varepsilon(W_1W_2:C\,|\,U)_\sigma \;-\; 2 -\epsilon_{5}
 \end{array}
\nonumber
\end{eqnarray}
All the above information quantities are computed with respect to the states
in \eqref{Eqn:TestChannelJointStates} and
\eqref{Eqn:TestChannelIndividualStates}. Define
$\ScrA_{\epsilon}(\rho^{A_1A_2C},\ulineCalW,\ulineCalI,p_{YU|\ulineW})$
as the set of all $(R_1,R_2,C_1,C_2)\in\reals_{\geq0}^{4}$ satisfying
 \begin{eqnarray}
 \label{Eqn:PreFM1}
 R_{1}\!> \sfB^{\epsilon}_{1}-\sfB^{\epsilon}_{6},~\!\!R_{2}\! > \sfB^{\epsilon}_{2}-\sfB^{\epsilon}_{7},~\!\!R_{1}\!+R_{2}\! > \sfB^{\epsilon}_{1}+\sfB^{\epsilon}_{2}-\sfB^{\epsilon}_{8},
 \!\!\!\!\!\\
 \label{Eqn:PreFM2}
 R_{1}+C_{1}  > \sfB^{\epsilon}_{3}-\sfB^{\epsilon}_{6},~R_{1}+R_{2}+C_{1} >\sfB^{\epsilon}_{2}+\sfB^{\epsilon}_{3}-\sfB^{\epsilon}_{8},\!\!\!
 \\
 \label{Eqn:PreFM3}
 R_{2}+C_{2} >\sfB^{\epsilon}_{4}-\sfB^{\epsilon}_{7},~R_{1}+R_{2}+C_{2}  >\sfB^{\epsilon}_{1}+\sfB^{\epsilon}_{4}-\sfB^{\epsilon}_{8},\!\!\!
 \\
 \label{Eqn:PreFM4}
 R_{1}+C_{1}+R_{2}+C_{2} > \max\{\sfB^{\epsilon}_{3}+\sfB^{\epsilon}_{4},\sfB^{\epsilon}_{5}\}-\sfB^{\epsilon}_{8}.\!\!\!
\end{eqnarray}
\end{fulldefinition}
We are now set to characterize a one-shot inner bound for multi-party instrument simulation.

\begin{fulltheorem}
\label{Thm:DistInstSimltnOneShot}
For $\eta>0$, instrument $\CalJ$'s action on $\rho^{A_{1}A_{2}C}$ can be $\eta$-simulated in one-shot with communication cost $(R_{1},R_{2},C_{1},C_{2})$ if there exists a $(\ulineCalW,\ulineCalI,p_{YU|\ulineW}) \in \ScrT(\CalJ,\rho^{A_{1}A_{2}})$ for which $(R_{1},R_{2},C_{1},C_{2}) \in \ScrA_{\tilde{\eta}}(\rho^{A_{1}A_{2}C},\allowbreak \ulineCalW,\ulineCalI,p_{YU|\ulineW})$ where $\tilde{\eta}=\lambda \eta^{6}$ and $\lambda>0$ is a fixed constant (given explicitly in the proof).
\end{fulltheorem}

\begin{proof}
 Fix a generic triple $(\ulineCalW,\ulineCalI,p_{YU|\ulineW}) \in \ScrT(\CalJ,\rho^{A_{1}A_{2}})$ and $\mathcal{I}^{(i)}(\rho_{A_i})
= \Tr_{E_i}\!\left[
\sum_{w_i\in\CalW_i}\
V_{w_i}\,\rho_{A_i}\,{V_{w_i}}^\dagger
\otimes \ketbra{w_i}_{W_{i}}
\right],$ where $V_{w_i}:\CalH_{A_i}\to \CalH_{B_i}\otimes \CalH_{E_i}$, $i\in\{1,2\}$, satisfy $\sum_{w_i\in\CalW_i}\ {V_{w_i}}^\dagger V_{w_i} = I_{A_i}$. Let $p_{W_1W_2}(w_1,w_2)\ \define\ \Tr\!\big[(\mu_{w_1}\otimes \mu_{w_2})\,\rho_{A_1A_2}\big],\mu_{w_i}\ \define\ {V_{w_i}}^\dagger V_{w_i}$. We begin by describing the $K_{1},K_{2}$ encoder instruments and the stochastic decoder.
\med\textit{\underline{The $K_{i}$ encoder instruments}}: Throughout, $K_{i}=2^{C_{i}},M_{i}=2^{R_{i}},B_{i}=2^{\beta_{i}} \in \naturals$ and $[K_{i}],[M_{i}],[B_{i}]$ denote common randomness, message and bin index sets respectively. Since the outcomes of the two instruments are correlated with the side information $C$, a binning based on one-shot quantum joint typicality \cite{202103SAD_Sen} can reduce message rates $R_{1},R_{2}$. The outcomes are therefore binned and the bin index $b_{i} \in [B_{i}]$ is not communicated to the Rx. Specifically, for $i \in [2]$, let $\sfc^{i} = (c_{k_{i}} : k_{i} \in [K_{i}])$ denote a multi-code comprising of $K_{i}=2^{C_{i}}$ codes wherein the $k_{i}-$th code $c_{k_{i}}=(w_{i}(k_{i},m_{i},b_{i}) \in \CalW_{i}:(m_{i},b_{i}) \in [M_{i}] \times [B_{i}])$ comprises of $M_{i}=2^{R_{i}}$ bins each with $B_{i}=2^{\beta_{i}}$ codewords. Associated with each codeword, $w_{i}(k_{i},m_{i},b_{i})$, let $V_{k_i,m_i,b_i} \define V_{w_i(k_i,m_i,b_i)}$ $\mu_{k_{i},m_{i},b_{i}} \define \mu_{w_{i}(k_{i},m_{i},b_{i})} \in \CalP(\CalH_{i})$.
Define the operators
\[
\begin{aligned}
\label{Eqn:DistLikelihoodPOVMdefn_i}
S_{k_i}
\define \sum_{m_i=1}^{M_i} \sum_{b_i=1}^{B_i}
\frac{\sqrt{\rho_{A_i}}\ \mu_{k_i,m_i,b_i}\ \sqrt{\rho_{A_i}}}
{M_i B_i\,\Tr(\rho_{A_i}\mu_{k_i,m_i,b_i})},~
z_{k_i}
\define \Pi_{\supp(S_{k_i})},
\end{aligned}
\]
and $\pi_{k_i}\define I_{A_i}-z_{k_i}$. For $m_i\in[M_i]$ define
\begin{equation}
\label{Eqn:DistLikelihoodTheta_i}
\theta^{m_i,b_i}_{k_i}
\define
\frac{
V_{k_i,m_i,b_i}\ \sqrt{\rho_{A_i}}\ {S_{k_i}}^{-\frac12}
}{
\sqrt{M_i B_i\,\Tr(\rho_{A_i}\mu_{k_i,m_i,b_i})}
},
\end{equation}
where ${S_{k_i}}^{-\frac12}$ denotes the generalized inverse (equal to $0$ on the null space of $S^{(i)}_{k_i}$). Note that, by definition $\theta^{m_i,b_i}_{k_i}: \CalH_{A_i}\to \CalH_{B_i}\otimes \CalH_{E_i} $.
Since $\pi_{k_i}$ is a projector, fix an isometry
$Q_i:\CalH_{A_i}\to \CalH_{B_i}\otimes\CalH_{E_i}$
and define the ``$0$-th'' operator
\[
\theta^{0}_{k_i}\ \define\ Q_i\,\pi_{k_i},
\qquad\text{so that}\qquad
{\theta^{0}_{k_i}}^\dagger \theta^{0}_{k_i}=\pi_{k_i}.
\]
Note that for every $(k_i,m_i,b_i)\in[K_i]\times[M_i]\times[B_i]$, we have
\begin{equation}
\label{Eqn:DistThetak0GetsKilled_i}
\theta^{0}_{k_i}\sqrt{S_{k_i}} = 0, \theta^{m_i, b_i}_{k_i}\sqrt{S_{k_i}}  =
\frac{
V_{k_i,m_i,b_i}\sqrt{\rho_{A_i}}
}{
\sqrt{M_i B_i\,\Tr(\rho_{A_i}\mu_{k_i,m_i,b_i})}
}.\nonumber
\end{equation}
It is also easy to verify that for each $(k_{i},m_{i},b_{i}) \in [K_{i}]\times [M_{i}] \times [B_{i}]$, $\theta^{m_i,b_i}_{k_i}: \CalH_{A_i} \to \CalH_{B_i}\otimes \CalH_{E_i} $ and ${\theta_{k_{i}}^{0}}^{\dagger}\theta_{k_{i}}^{0}+\sum_{m_{i},b_{i}}{\theta_{k_{i}}^{m_{i},b_{i}}}^{\dagger}\theta_{k_{i}}^{m_{i},b_{i}} = I_{A_{i}}$
We are now set to define the two banks of encoding instruments. For $ i \in [2]$, Alice$_{i}$'s instrument corresponding to common randomness index $k_{i} \in [K_{i}]$ is $\CalE_{k_{i}}^{(i)}:\ \CalL(\CalH_{A_i})\longrightarrow \CalL(\CalH_{B_i}\otimes \mathbb{C}^{M_i})$, defined through its action
\begin{eqnarray}
 \label{Eqn:EncodingInstrumentki}
 \CalE_{k_{i}}^{(i)}(S)= \tr_{E_{i}}\!\left[\!\!\!\begin{array}{r}\displaystyle\sum_{m_i=1}^{M_i} \sum_{b_i=1}^{B_i}
\theta^{m_i, b_i}_{k_i}\,S\,
{\theta^{m_i, b_i}_{k_i}}^\dagger\!\!\!
\otimes \ketbra{m_i}_{M_i}
+
\theta^{0}_{k_i}\,S\,
{\theta^{0}_{k_i}}^\dagger
\otimes \ketbra{0}_{M_i} \end{array}\!\!\!\right]
\end{eqnarray}
on any $S \in \CalL(\CalH_{A_{i}})$.
\begin{fullremark}
 \label{Rem:InverseOfRandomVariables}
The inverse $S_{k_i}^{-1/2}$ in \eqref{Eqn:DistLikelihoodTheta_i} is
codebook-dependent, which prevents a direct concentration argument.
This is the reason for introducing the proxy states below.
\end{fullremark}

\med\textit{\underline{$K_{1}K_{2}$ Decoder post-processing maps}}: Recall that the bin indices are not communicated to Charlie. The decoder $\CalD_{\ulinek} : \CalL(\CalH_{C}\otimes\mathbb{C}^{M_1}\otimes \mathbb{C}^{M_2})
\longrightarrow \CalL(\CalH_{C}\otimes\mathbb{C}^{|\mathcal{Y}|})$ is therefore a composition $\CalD_{\ulinek} \define \CalD_{\ulinek}^{(2)} \circ \CalD_{\ulinek}^{(1)} $, wherein the $\CalD_{\ulinek}^{(1)}:\CalL(\CalH_{C}\otimes\mathbb{C}^{\ulineM}) \rightarrow \CalL(\CalH_{C}\otimes\mathbb{C}^{\ulineM} \otimes \mathbb{C}^{\ulineB})$ is a hypothesis-testing-based decoder used in CQMAC decoding \cite{202103SAD_Sen} to recover the bin index and $\CalD_{\ulinek}^{(2)}$ is a plain stochastic processing defined as
\begin{equation}
  \CalD_{\ulinek}^{(2)}(S)= \sum_{\ulinem \in [\ulineM]}\sum_{y \in \CalY}p_{Y|\ulineW}(y|\ulinew(\ulinek,\ulinem,\ulineb))\ketbra{y}{\ulinem ~\ulineb}S\ketbra{\ulinem ~\ulineb}{y}
  \nonumber
\end{equation}
for any $S \in \CalL(\mathbb{C}^{\ulineM} \otimes \mathbb{C}^{\ulineB})$ where $\ulinew(\ulinek,\ulinem,\ulineb)$, $\mathbb{C}^{\ulineM}$, $\mathbb{C}^{\ulineB}$ abbreviate $w_{1}(k_{1},m_{1},b_{1}),w_{2}(k_{2},m_{2},b_{2})$, $\mathbb{C}^{M_1}\otimes \mathbb{C}^{M_2}$, $\mathbb{C}^{B_1}\otimes \mathbb{C}^{B_2}$ respectively. Regarding the former, one can consider that the realization of the instruments $\CalI^{(i)}$ on the state $\rho_{A_1A_2C}$ induces a CQMAC channel on Charlie's side, in the sense that his part of the system collapses to the state $\rho_C^{w_1,w_2}$ with (input) probability $p_{W_1W_2}(w_1,w_2)$. For each subcode $\big(\ulinew(\ulinek,\ulinem,\ulineb)\big)_{\ulineb \in [\ulineB]}$, we therefore associate a CQMAC code (provided in \cite{202103SAD_Sen}) specified by a POVM $ D^{(\ulinek,\ulinem)}=\{D_{\ulinek,\ulinem}^{\,\ulineb}\}_{\ulineb \in [\ulineB]}$ on $\mathcal H_C$, enabling Charlie to recover the bin index. More formally:
\begin{equation}
  \CalD_{\ulinek}^{(1)}(S)
  = \sum_{\ulineb, \ulinem}
  \Bigl(\sqrt{D_{\ulinek,\ulinem}^{\,\ulineb}}\otimes \ketbra{\ulinem ~\ulineb}{\ulinem}\Bigr)\,
  S\,
  \Bigl(\sqrt{D_{\ulinek,\ulinem}^{\,\ulineb}}\otimes \ketbra{\ulinem}{\ulinem ~\uline b}\Bigr)
  \nonumber
\end{equation}
\med\textit{{Analysis Outline}}:
We now sketch the proof and indicate where each family of inequalities in
Definition~\ref{Defn:OneShotPreFMInfoQuantities} enters.
Let
\begin{eqnarray}
\widetilde{\CalT}
=
\frac{1}{K_1K_2}
\sum_{\ulinek}
\left[\!\!\!\begin{array}{c}
(\mathrm{id}_{XB_1B_2}\otimes \CalD_{\ulinek})
\circ
(\mathrm{id}_{XC}\otimes \CalE_{k_1}^{(1)}\otimes \CalE_{k_2}^{(2)})
\end{array}\!\!\!\right]
\nonumber
\left\{\varphi^{XA_1A_2C}\right\}.
\nonumber
\end{eqnarray}
For $\ulinew=(w_1,w_2)$, define the conditional states
\begin{equation}
\label{Eqn:QSIConditionalTargetState}
\begin{gathered}
\CalT_{XB_1B_2C}^{\ulinew}
\define
\Tr_{E_1E_2}\!\left(
\frac{
(\mathrm{id}_{XC}\otimes V_{\ulinew})\,
\varphi^{XA_1A_2C}\,
(\mathrm{id}_{XC}\otimes V_{\ulinew}^{\dagger})
}{
p_{\ulineW}(\ulinew)
}
\right)\\
\CalT_{XB_1B_2CY}^{\ulinew}
\define
\sum_{y}
p_{Y|\ulineW}(y|\ulinew)\,
\CalT_{XB_1B_2C}^{\ulinew}\otimes \ketbra{y},
\end{gathered}
\end{equation}
so that
\(
\CalT=\sum_{\ulinew} p_{\ulineW}(\ulinew)\CalT_{XB_1B_2CY}^{\ulinew}.
\) We next introduce proxy states
$\CalT_1,\CalT_2,\CalT_3$
and split the simulation error into four contributions:
\begin{equation}
\label{Eqn:QSIErrorBreakdown}
\begin{aligned}
\norm{\CalT-\widetilde{\CalT}}_1
\le
\norm{\CalT-\CalT_1}_1
+
\norm{\CalT_1-\CalT_2}_1 
+
\norm{\CalT_2-\CalT_3}_1
+
\norm{\CalT_3-\widetilde{\CalT}}_1 .
\end{aligned}
\end{equation}
The four terms correspond respectively to:
distributed QC covering, CQMAC packing, and two quantum-only covering terms.

\smallskip
\noindent\textit{(i) QC covering.}
Define
\begin{eqnarray}
\CalT_1
=
\frac{1}{KMB}
\sum_{\ulinek,\ulinem,\ulineb}
\frac{
p_{\ulineW}(\ulinew(\ulinek,\ulinem,\ulineb))\CalT_{XB_1B_2CY}^{\,\ulinew(\ulinek,\ulinem,\ulineb)}
}{
p_{W_1}(w_1(k_1,m_1,b_1))\,
p_{W_2}(w_2(k_2,m_2,b_2))
}.
\nonumber
\end{eqnarray}
The first term directly reduces to a soft-covering estimate. Hence, by the multipartite one-shot covering bound, $\Expectation_{\mathcal C}\!\left[\norm{\CalT-\CalT_1}_1\right]
\le 16\epsilon$ provided that the rates satisfy $\forall i\in[2],$
\begin{eqnarray}
\label{Eqn:QSIQCOne}
R_i+C_i+\beta_i &>& I_2^\epsilon(W_i:YC\ulineB X)_\sigma+\delta,
\nonumber\\[-0.1em]
R_1+R_2+C_1+C_2+\beta_1+\beta_2
&>& I_\sigma^\epsilon+\delta
\nonumber
\end{eqnarray}
where $I_\sigma^\epsilon\define
D_2^\epsilon\!\left(
\sigma^{XC\ulineB\ulineW Y}\,\middle\|\,
\sigma^{W_1}\!\otimes\!\sigma^{W_2}\!\otimes\!\sigma^{XC\ulineB Y}
\right)_\sigma$ and $\delta=2\log\tfrac{1}{\epsilon}$. These are precisely the constraints producing
$\sfB_3^\epsilon,\sfB_4^\epsilon,\sfB_5^\epsilon$.

\smallskip
\noindent\textit{(ii) CQMAC packing term.}
The second proxy state \(\CalT_2\) is obtained from \(\CalT_1\) by applying,
for every \((\ulinek,\ulinem)\), the CQMAC POVM $D^{(\ulinek,\ulinem)}
=
\{D_{\ulinek,\ulinem}^{\,\widehat{\ulineb}}\}_{\widehat{\ulineb}\in[\ulineB]}$ on Charlie's system \(C\), and then using the reconstructed index
\(\widehat{\ulineb}\) in the stochastic post-processing
\(p_{Y|\ulineW}\). More explicitly, for $\ulinew_{\ulineb}
\define
\ulinew(\ulinek,\ulinem,\ulineb),$ and $a_{\ulinek,\ulinem,\ulineb}
\define
\frac{p_{\ulineW}(\ulinew_{\ulineb})}
{p_{W_1}(w_1(k_1,m_1,b_1))
 p_{W_2}(w_2(k_2,m_2,b_2))},$ we can write the corresponding term in \(\CalT_1\) as $\sum_{y}
p_{Y|\ulineW}(y|\ulinew_{\ulineb})
\CalT^{\ulinew_{\ulineb}}_{X\ulineB C}
\otimes \ketbra{y}_Y ,$ whereas the corresponding term in \(\CalT_2\) is $\sum_{\widehat{\ulineb}}
\sum_y
p_{Y|\ulineW}(y|\ulinew_{\widehat{\ulineb}})
\sqrt{D_{\ulinek,\ulinem}^{\,\widehat{\ulineb}}}\,
\CalT^{\ulinew_{\ulineb}}_{X\ulineB C}\,
\sqrt{D_{\ulinek,\ulinem}^{\,\widehat{\ulineb}}}
\otimes\ketbra{y}_Y ,$ where the POVM element acts on \(C\) and the identity on \(X\ulineB\) is
suppressed. We split the last expression according to whether
\(\widehat{\ulineb}=\ulineb\) or \(\widehat{\ulineb}\neq\ulineb\). By the
triangle inequality,
\begin{eqnarray}
\norm{\CalT_1-\CalT_2}_1
&\leq&
\frac{1}{KMB}
\sum_{\ulinek,\ulinem,\ulineb}
a_{\ulinek,\ulinem,\ulineb}
\Bigl(
\Delta^{\mathrm{g}}_{\ulinek,\ulinem,\ulineb}
+
\Delta^{\mathrm{b}}_{\ulinek,\ulinem,\ulineb}
\Bigr),
\label{Eqn:CQMACGoodBadSplit}
\nonumber
\end{eqnarray}
where the good-output disturbance term is
\[
\Delta^{\mathrm{g}}_{\ulinek,\ulinem,\ulineb}
\define
\left\|
\CalT^{\ulinew_{\ulineb}}_{X\ulineB C}
-
\sqrt{D_{\ulinek,\ulinem}^{\,\ulineb}}\,
\CalT^{\ulinew_{\ulineb}}_{X\ulineB C}\,
\sqrt{D_{\ulinek,\ulinem}^{\,\ulineb}}
\right\|_1
\]
and the bad-output term \(\Delta^{\mathrm{b}}_{\ulinek,\ulinem,\ulineb}\)
is defined by the following trace norm:
\[
\left\|
\sum_{y,\widehat{\ulineb}\neq\ulineb}
p_{Y|\ulineW}(y|\ulinew_{\widehat{\ulineb}})
\sqrt{D_{\ulinek,\ulinem}^{\,\widehat{\ulineb}}}\,
\CalT^{\ulinew_{\ulineb}}_{X\ulineB C}\,
\sqrt{D_{\ulinek,\ulinem}^{\,\widehat{\ulineb}}}
\otimes\ketbra{y}_Y
\right\|_1 .
\]
The second term is positive, hence its trace norm is its trace. Therefore, $\Delta^{\mathrm{b}}_{\ulinek,\ulinem,\ulineb}
=
\sum_{\widehat{\ulineb}\neq\ulineb}
\tr\!\left[
D_{\ulinek,\ulinem}^{\,\widehat{\ulineb}}\,
\CalT^{\ulinew_{\ulineb}}_{C}
\right]
=
1-
\tr\!\left[
D_{\ulinek,\ulinem}^{\,\ulineb}\,
\CalT^{\ulinew_{\ulineb}}_{C}
\right].$ On the other hand, the first term measures
the disturbance induced on the side-information system \(C\) when the correct
outcome \(\ulineb\) is obtained. By the gentle operator lemma, $\Delta^{\mathrm{g}}_{\ulinek,\ulinem,\ulineb}
\leq
2
\sqrt{
1-
\tr\!\left[
D_{\ulinek,\ulinem}^{\,\ulineb}\,
\CalT^{\ulinew_{\ulineb}}_{C}
\right]
}.$ Define the weighted CQMAC decoding error
\[
P_{\mathrm{err}}(\CalC)
\define
\frac{1}{KMB}
\sum_{\ulinek,\ulinem,\ulineb}
a_{\ulinek,\ulinem,\ulineb}
\left(
1-
\tr\!\left[
D_{\ulinek,\ulinem}^{\,\ulineb}\,
\CalT^{\ulinew_{\ulineb}}_{C}
\right]
\right).
\]
Combining the previous estimates gives
\begin{eqnarray}
\norm{\CalT_1-\CalT_2}_1
\leq
P_{\mathrm{err}}(\CalC)
+
\frac{2}{KMB}
\sum_{\ulinek,\ulinem,\ulineb}
\left[\!\!\!\begin{array}{c}
a_{\ulinek,\ulinem,\ulineb}
\sqrt{
1-
\tr\!\left[
D_{\ulinek,\ulinem}^{\,\ulineb}\,
\CalT^{\ulinew_{\ulineb}}_{C}
\right]
}
\end{array}\!\!\!\right].
\label{Eqn:CQMACBeforeJensen}
\nonumber
\end{eqnarray}

Taking expectation over the random codebook and using Cauchy--Schwarz,
together with $\Expectation_{\CalC}
\left[
\frac{1}{KMB}
\sum_{\ulinek,\ulinem,\ulineb}
a_{\ulinek,\ulinem,\ulineb}
\right]
=1,$ we obtain $\Expectation_{\CalC}
\left[
\norm{\CalT_1-\CalT_2}_1
\right]
\leq
\Expectation_{\CalC}\!\left[P_{\mathrm{err}}(\CalC)\right]
+
2
\sqrt{
\Expectation_{\CalC}\!\left[P_{\mathrm{err}}(\CalC)\right]
}.$ The likelihood factor
$a_{\ulinek,\ulinem,\ulineb}$ changes measure: after averaging over the
codebook, it converts the product law used to generate
$(W_1(k_1,m_1,b_1),W_2(k_2,m_2,b_2))$ into $p_{W_1W_2}$. Using
$p_{W_1W_2}=\sum_u p_U(u)p_{W_1|U}(w_1|u)p_{W_2|U}(w_2|u)$, the weighted
error $P_{\mathrm{err}}(\CalC)$ is an average over $U$ of conditional
CQMAC error criteria. Thus, for every fixed $(\ulinek,\ulinem)$, the
subcodebook $\bigl(\ulinew(\ulinek,\ulinem,\ulineb)\bigr)_{\ulineb\in[\ulineB]}$
induces a CQMAC with output system $C$, and Sen's one-shot simultaneous
decoding theorem, applied conditionally on $U$, gives
$\Expectation_{\CalC}\!\left[P_{\mathrm{err}}(\CalC)\right]
\leq c_0\epsilon$ for a universal constant $c_0>0$, provided that $\forall i\in[2],$
\begin{eqnarray}
\label{Eqn:QSICQMACPacking}
\beta_i
&<&
I_H^\epsilon(W_i:W_{3-i}C|U)_\sigma-2-\log\tfrac{1}{\epsilon},
\nonumber\\[-0.1em]
\beta_1+\beta_2
&<&
I_H^\epsilon(W_1W_2:C|U)_\sigma-2-\log\tfrac{1}{\epsilon}.
\end{eqnarray}
Substituting this estimate yields $\Expectation_{\CalC}
\left[
\norm{\CalT_1-\CalT_2}_1
\right]
\leq
c\sqrt{\epsilon}$ for another universal constant \(c>0\). These are exactly the packing
inequalities encoded by
\(\sfB_6^\epsilon,\sfB_7^\epsilon,\sfB_8^\epsilon\).

\smallskip
\noindent\textit{(iii)--(iv) Quantum-only covering and compatible-operator-sliding terms.}
We now spell out the two remaining estimates. These terms are quantum-only in the sense that they do not involve the stochastic post-processing error or the CQMAC decoding error. Their role is to remove the random inverse operators appearing in the likelihood instruments while preserving the post-instrument systems. To avoid a clash with the classical random variables $W_1,W_2$, let $\mathsf{V}_i$ denote an isometry from the purifying space of $\varphi_{\rho_{A_i}}$ into $XCA_{3-i}$ such that
\begin{equation}
\label{Eqn:QSIIsometricExtensionState}
\varphi^{XA_1A_2C}
=
(\mathsf{V}_i\otimes I_{A_i})\,
\varphi_{\rho_{A_i}}\,
(\mathsf{V}_i^*\otimes I_{A_i}),
\qquad i\in[2].
\end{equation}
Define the final proxy state by replacing, on Alice$_1$'s side, the true input purification by the canonical purification associated with the likelihood normalizer $S_{k_1}$:
\begin{eqnarray}
\label{Eqn:QSIT3BeforeCOS}
\CalT_3
=
\frac{1}{K_1K_2}
\sum_{\ulinek}
\left[\!\!\!\begin{array}{c}
(\mathrm{id}_{XB_1B_2}\otimes \CalD_{\ulinek})
\circ
(\mathrm{id}_{XC}\otimes \CalE_{k_1}^{(1)}\otimes \CalE_{k_2}^{(2)})
\end{array}\!\!\!\right]
\left(
(\mathsf{V}_1\otimes I_{A_1})\,
\varphi_{S_{k_1}}\,
(\mathsf{V}_1^*\otimes I_{A_1})
\right).
\end{eqnarray}
Since $\widetilde{\CalT}$ is obtained from the same CPTP maps with $\varphi^{XA_1A_2C}$ in place of the displayed input, contractivity of the trace norm gives
\begin{eqnarray}
\label{Eqn:QSIT3TildeExpansion}
\norm{\CalT_3-\widetilde{\CalT}}_1
&\leq&
\frac{1}{K_1K_2}
\sum_{\ulinek}
\left\|
(\mathsf{V}_1\otimes I_{A_1})\,
\varphi_{S_{k_1}}\,
(\mathsf{V}_1^*\otimes I_{A_1})
-
\varphi^{XA_1A_2C}
\right\|_1
\nonumber\\
&=&
\frac{1}{K_1}
\sum_{k_1=1}^{K_1}
\norm{\varphi_{S_{k_1}}-\varphi_{\rho_{A_1}}}_1.
\end{eqnarray}
Using Lemma~\ref{lem:canonical-purif-stability}, Jensen's inequality for the concave fourth-root function, and the one-party instance of Proposition~\ref{Prop:SmoothMultipartiteSoftCovering}, we obtain
\begin{eqnarray}
\label{Eqn:QSIT3TildeBound}
\Expectation_{\CalC}\!\bigl[\norm{\CalT_3-\widetilde{\CalT}}_1\bigr]
&\leq&
2\sqrt{2}\,
\Expectation_{\CalC}\!\left[
\frac{1}{K_1}
\sum_{k_1=1}^{K_1}
\sqrt[4]{\norm{S_{k_1}-\rho_{A_1}}_1}
\right]
\nonumber\\
&\leq&
2\sqrt{2}\,
\sqrt[4]{
\Expectation_{\CalC}\!\left[
\left\|
\frac{1}{M_1B_1}
\sum_{m_1,b_1}
\frac{\sqrt{\rho_{A_1}}\mu_{k_1,m_1,b_1}\sqrt{\rho_{A_1}}}
{p_{W_1}(w_1(k_1,m_1,b_1))}
-
\rho_{A_1}
\right\|_1
\right]
}
\nonumber\\
&\leq&
2\sqrt{2}\,\epsilon^{1/4},
\end{eqnarray}
provided that
\begin{equation}
\label{Eqn:QSIQuantumCovering1}
R_1+\beta_1
>
I_2^\epsilon(W_1:CA_2X)_{\sigma_1}
+2\log\frac{1}{\epsilon}.
\end{equation}
This is the constraint encoded by $\sfB_1^\epsilon$.

It remains to bound $\norm{\CalT_2-\CalT_3}_1$. This is the only place where the random inverse $S_{k_i}^{-1/2}$ in the likelihood instruments must be handled carefully. We introduce the inverse-free maps
\begin{eqnarray}
\label{Eqn:QSIAuxiliaryMaps}
\widetilde{\CalI}_{k_i}^{(i)}(S)
=
\Tr_{E_i}\!\left[\!\!\!\begin{array}{r}
\displaystyle\sum_{m_i=1}^{M_i}\sum_{b_i=1}^{B_i}
\frac{
V_{w_i(k_i,m_i,b_i)}\,S\,
V_{w_i(k_i,m_i,b_i)}^\dagger
}{
M_iB_i\,p_{W_i}(w_i(k_i,m_i,b_i))
}
\displaystyle\otimes \ketbra{m_i}_{M_i}
\end{array}\!\!\!\right].
\end{eqnarray}
The compatible-operator-sliding computation in Appendix~\ref{AppSec:QSI-COS} gives the two equivalent representations
\begin{eqnarray}
\label{Eqn:QSICOST2After}
\CalT_2
&=&
\frac{1}{K_1K_2}
\sum_{\ulinek}
\left[\!\!\!\begin{array}{c}
(\mathrm{id}_{XB_1B_2}\otimes \CalD_{\ulinek})
\circ
(\mathrm{id}_{XC}\otimes \widetilde{\CalI}_{k_1}^{(1)}\otimes \CalE_{k_2}^{(2)})
\end{array}\!\!\!\right]
\left(
(\mathsf{V}_2\otimes I_{A_2})\,
\varphi_{S_{k_2}}\,
(\mathsf{V}_2^*\otimes I_{A_2})
\right),
\end{eqnarray}
where the decoder is the same hypothesis-testing decoder followed by stochastic post-processing as in the definition of $\CalT_2$, and
\begin{eqnarray}
\label{Eqn:QSICOST3After}
\CalT_3
&=&
\frac{1}{K_1K_2}
\sum_{\ulinek}
\left[\!\!\!\begin{array}{c}
(\mathrm{id}_{XB_1B_2}\otimes \CalD_{\ulinek})
\circ
(\mathrm{id}_{XC}\otimes \widetilde{\CalI}_{k_1}^{(1)}\otimes \CalE_{k_2}^{(2)})
\end{array}\!\!\!\right]
\left(\varphi^{XA_1A_2C}\right).
\end{eqnarray}
The point of \eqref{Eqn:QSICOST2After}--\eqref{Eqn:QSICOST3After} is that both states are now acted on by the same maps; only the input on Alice$_2$'s side differs. Moreover, Lemma~\ref{Lem:QSIInverseFreeMapContraction} shows that the averaged inverse-free map is trace-norm contractive, even with an arbitrary reference system. Therefore, applying first contractivity of the decoder and of $\CalE_{k_2}^{(2)}$, then Lemma~\ref{Lem:QSIInverseFreeMapContraction}, and finally Lemma~\ref{lem:canonical-purif-stability}, yields
\begin{eqnarray}
\label{Eqn:QSIT2T3Bound}
\Expectation_{\CalC}\!\left[\norm{\CalT_2-\CalT_3}_1\right]
&\leq&
\frac{1}{K_2}
\sum_{k_2=1}^{K_2}
\Expectation_{\CalC}\!\left[
\left\|
(\mathsf{V}_2\otimes I_{A_2})\,
\varphi_{S_{k_2}}\,
(\mathsf{V}_2^*\otimes I_{A_2})
-
\varphi^{XA_1A_2C}
\right\|_1
\right]
\nonumber\\
&=&
\frac{1}{K_2}
\sum_{k_2=1}^{K_2}
\Expectation_{\CalC}\!\left[
\norm{\varphi_{S_{k_2}}-\varphi_{\rho_{A_2}}}_1
\right]
\nonumber\\
&\leq&
2\sqrt{2}\,
\sqrt[4]{
\Expectation_{\CalC}\!\left[
\left\|
\frac{1}{M_2B_2}
\sum_{m_2,b_2}
\frac{\sqrt{\rho_{A_2}}\mu_{k_2,m_2,b_2}\sqrt{\rho_{A_2}}}
{p_{W_2}(w_2(k_2,m_2,b_2))}
-
\rho_{A_2}
\right\|_1
\right]
}
\nonumber\\
&\leq&
2\sqrt{2}\,\epsilon^{1/4},
\end{eqnarray}
provided that
\begin{equation}
\label{Eqn:QSIQuantumCovering2}
R_2+\beta_2
>
I_2^\epsilon(W_2:CA_1X)_{\sigma_2}
+2\log\frac{1}{\epsilon}.
\end{equation}
This is the constraint encoded by $\sfB_2^\epsilon$. Hence the two quantum-only terms are controlled by the single-sender covering constraints $\sfB_1^\epsilon$ and $\sfB_2^\epsilon$, while Appendix~\ref{AppSec:QSI-COS} justifies the change of representation needed to remove the likelihood inverses.

Collecting the four bounds above and applying Fourier--Motzkin elimination to
the auxiliary binning rates $(\beta_1,\beta_2)$ yields exactly the region in
Definition~\ref{Defn:OneShotPreFMInfoQuantities}.
Absorbing the universal constants into a single constant $\lambda>0$ and
choosing $\epsilon=\tilde\eta=\lambda\eta^6$ concludes the proof.
\end{proof}
\subsection{Instrument Simulation via Coset Codes : Asymptotic Regime}
\label{SubSec:InstSimltnAsymptoticCoset}
We now state an asymptotic achievability result based on coset-structured
codes. This construction differs from the IID-code construction: the
decoder exploits the finite-field structure to recover the sum of the hidden
indices. We refer to \cite{202403arXiv_Pad} for details on Sec.~\ref{SubSec:InstSimltnAsymptoticCoset}.

Let $\mathbb{F}_{q}$ be a finite field, with addition denoted by $\oplus$,
and set $\CalW_{1}=\CalW_{2}=\mathbb{F}_{q}$. Let
$W_{\oplus}\define W_{1}\oplus W_{2}$, and let $\pi_q$ denote the uniform
distribution on $\mathbb{F}_{q}$. For an instrument $\CalJ$ and state
$\rho^{A_1A_2C}$, let
$\ScrT_{\oplus}(\CalJ,\rho^{A_1A_2C})$ denote the collection of all triples
$(\ulineCalW,\ulineCalI,p_{Y|W_{\oplus}})$ such that
$\CalI^{(i)}:L(\CalH_{A_i})\to L(\CalH_{B_i}\otimes\mathbb{C}^{\CalW_i})$
is an instrument for $i\in[2]$, and, for every $y\in\CalY$, $\CalJ_y
=
\sum_{w_1,w_2\in\mathbb{F}_{q}}
p_{Y|W_{\oplus}}(y|w_1\oplus w_2)\,
\CalI^{(1)}_{w_1}\otimes \CalI^{(2)}_{w_2}.$ For such a triple, define $\sigma^{XC\ulineB\ulineW W_{\oplus}Y}
\define
\left(
\mathrm{id}_{XC}\otimes
\olinescre^{Y W_{\oplus}|\ulineW}_{p}
\otimes \CalI
\right)
\bigl(\varphi^{XA_1A_2C}\bigr),$ where $\CalI=\CalI^{(1)}\otimes\CalI^{(2)}$ and $p_{Y W_{\oplus}|\ulineW}(y,w|w_1,w_2)
=
\mathbf{1}\{w=w_1\oplus w_2\}\,
p_{Y|W_{\oplus}}(y|w).$ Also define, for $i\in[2]$, $\sigma_i^{XB_iA_{3-i}CW_i}
\define
\left(
\mathrm{id}_{XA_{3-i}C}\otimes \CalI^{(i)}
\right)
\bigl(\varphi^{XA_1A_2C}\bigr).$ All information quantities below are computed with respect to these states. Set $\Delta_i \define D(p_{W_i}\|\pi_q)$ and $
\Gamma_{\oplus}\define I(W_{\oplus}:C)_{\sigma}
+D(p_{W_{\oplus}}\|\pi_q).$
\begin{fulltheorem}[Asymptotic instrument simulation via coset codes]
\label{Thm:AsymptoticCosetInstSimulation}
Instrument $\CalJ$'s action on $\rho^{A_1A_2C}$ can be perfectly simulated
with communication and common-randomness cost
$(R_1,R_2,C_1,C_2)$ if there exists
$(\ulineCalW,\ulineCalI,p_{Y|W_{\oplus}})
\in \ScrT_{\oplus}(\CalJ,\rho^{A_1A_2C})$
such that $\forall i\in[2],$
\begin{eqnarray}
R_i
&>&
I(W_i:CA_{3-i}X)_{\sigma_i}
+\Delta_i-\Gamma_{\oplus},
\nonumber\\[-0.1em]
R_i+C_i
&>&
I(W_i:YC\ulineB X)_{\sigma}
+\Delta_i-\Gamma_{\oplus},
\nonumber\\[-0.1em]
R_1+R_2+C_1+C_2
&>&
I(\ulineW:YC\ulineB X)_{\sigma}
+
I(W_1:W_2)_{\sigma}
\nonumber
+\Delta_1+\Delta_2
-2\Gamma_{\oplus}.
\label{Eqn:CosetAsymptoticRegion}
\end{eqnarray}
\end{fulltheorem}

We first isolate the point-to-point measurement-compression component based on coset-structured likelihood POVMs. This component has no receiver quantum side information and concerns a single POVM $\lambda_{\CalY}$ acting on a single system with state $\rho$. Its role is only to provide the algebraic likelihood-POVM block used later in the distributed construction.

To obtain Theorem~\ref{Thm:AsymptoticCosetInstSimulation}, this point-to-point coset block is combined with the distributed binning and decoding mechanism used in the proof of Theorem~\ref{Thm:DistInstSimltnOneShot}. Each transmitter uses the corresponding local structured likelihood instrument, the hidden bin indices are not communicated, and Charlie uses the QSI system $C$ together with a hypothesis-testing based classical--quantum decoder to recover the relevant algebraic component. The packing constraints come from this decoding step, while the covering and proxy-state terms are inherited from the point-to-point coset-code measurement-compression argument recalled below.
\begin{proof}
The next paragraphs temporarily use the notation of a point-to-point POVM simulation problem. Thus $\rho\in\CalD(\CalH)$ is a single input state, $X$ is its inaccessible reference, $\lambda_{\CalY}$ is the target POVM, and $(\CalW,\mu_{\CalW},p_{Y|W})$ is a point-to-point test triple. We identify $[K],[M]$ and $[B]$ with $\fieldq^{\kappa},\fieldq^r$ and $\fieldq^{\beta}$, respectively, so that $K=q^{\kappa}$, $M=q^r$ and $B=q^{\beta}$.
\med
\textit{Preliminaries and simulation POVM.}
Fix a point-to-point test triple $(\CalW,\mu_{\CalW},p_{Y|W})$ satisfying
$C+R>I(XY;W)_\sigma$ and $R>I(X;W)_\sigma$, where
$\sigma_{XWY}\define(i_X\otimes\overline{\mathscr E}^{Y|W}_p)\circ(i_X\otimes\mathscr E^{\mu})(\ketbra{\phi_\rho}{\phi_\rho})$.
Our goal in this auxiliary point-to-point block is to specify an $(n,C,R,\theta_{\mathrm{ptp}},\Delta_{\mathrm{ptp}})$ simulation protocol for which $\norm{\alphaorig-\alphasim}_{1}$ shrinks exponentially in $n$. The original and simulated states are
\begin{eqnarray}
 \label{Eqn:AlphaOrigDefnAgain}
 \alphaorig \define (i_{X}^{ n}\!\otimes \mathscr{E}^{\lambda^{n}}\!)(\ketbra{\phi_{\rho^{\otimes n}}}{\phi_{\rho^{\otimes n}}} ),  \alphasim \define  \SimulationCPTP(\ketbra{\phi_{\rho_{\ulineK}^{\otimes n}}\!\!}{\!\phi_{\rho_{\ulineK}^{\otimes n}}\!} ),\!\!\!\\
 \label{Eqn:AlphaSimDefnAgain}
 \SimulationCPTP\define (i_{X}^{n}\!\otimes\!\tr_{\ulineK}\!\otimes i_{\CalY^{n}})\circ(i_{X}^{ n}\!\otimes\! i_{\mathcal{H}_{\ulineK}}\!\otimes\! \mathscr{E}^{\Delta_{\mathrm{ptp}}}\! )\!\circ\! (i_{X}^{ n}\!\otimes\! i_{\mathcal{H}_{\ulineK}}\!\otimes\!  \mathscr{E}^{\theta_{\mathrm{ptp}}}).\!\!\!
\end{eqnarray}
We enlarge $\CalW$ to the finite field $\fieldq$ by adding dummy symbols, if needed. Each dummy symbol is assigned the zero POVM operator, and the corresponding rows of $p_{Y|W}$ are arbitrary. Hence we may assume throughout the point-to-point block that $\CalW=\fieldq$ and $\mu_{\CalW}=\{\mu_w\in\CalP(\CalH):w\in\CalW\}$.

Let $\omega \define \rho^{\otimes n}$. Let $g_c \in \CalW^{\kappa\times n}$, $g_m \in \CalW^{r\times n}$ and $g_b \in \CalW^{\beta\times n}$ be generator matrices and let $d^n\in\CalW^n$ be a bias vector. Define
$w^n(\ulinek,\ulinem,\ulineb)\define c(\ulinek,\ulinem,\ulineb)\define \ulinek g_c\oplus \ulinem g_m\oplus \ulineb g_b\oplus d^n$.
For this word, set $\mu_{\ulinek,\ulinem,\ulineb}\define\bigotimes_{t=1}^n\mu_{w(\ulinek,\ulinem,\ulineb)_t}$ and
$\upsilon(\ulinek,\ulinem,\ulineb)\define q^n\tr(\mu_{\ulinek,\ulinem,\ulineb}\omega)$.
The point-to-point likelihood POVM $\theta_{\mathrm{ptp}}$ is defined by
\begin{eqnarray}
\label{Eqn:PTPCosetLikelihoodPOVM}
\theta_{\ulinek,\ulinem}
&\define&
\sum_{\ulineb\in\CalW^\beta}\theta_{\ulinek,\ulinem,\ulineb},
\nonumber\\[-0.2em]
\theta_{\ulinek,\ulinem,\ulineb}
&\define&
\upsilon(\ulinek,\ulinem,\ulineb)
\frac{S_{\ulinek}^{-1/2}\sqrt{\omega}\mu_{\ulinek,\ulinem,\ulineb}\sqrt{\omega}S_{\ulinek}^{-1/2}}
{\tr(\omega\mu_{\ulinek,\ulinem,\ulineb})},
\nonumber\\[-0.2em]
S_{\ulinek}
&\define&
\sum_{\ulinem\in\CalW^r}\sum_{\ulineb\in\CalW^\beta}
\upsilon(\ulinek,\ulinem,\ulineb)
\frac{\sqrt{\omega}\mu_{\ulinek,\ulinem,\ulineb}\sqrt{\omega}}
{\tr(\omega\mu_{\ulinek,\ulinem,\ulineb})},
\qquad
\theta_{\ulinek,0}\define I_{\CalH}^{\otimes n}-z_{\ulinek},
\end{eqnarray}
where $z_{\ulinek}$ is the projector onto the support of $S_{\ulinek}$ and the convention $0/0=0$ is used. Then
$\theta_{\mathrm{ptp}}\define\{\overline{\theta}_{\ulinek,\ulinem}\define\theta_{\ulinek,\ulinem}\otimes\ketbra{\ulinek}:(\ulinek,\ulinem)\in\CalW^\kappa\times\CalW^r\}\cup\{\theta_{\ulinek,0}\otimes\ketbra{\ulinek}\}$ is a valid POVM by the usual support-projection argument.
For every $(\ulinek,\ulinem) \in \CalW^{\kappa} \times \CalW^{r}$, we let 
\begin{eqnarray}
\label{Eqn:NoTypicalElements}
\mathrm{l}(\ulinek,\ulinem) =\sum_{\ulineb \in \CalW^{\beta}}\mathds{1}_{\{ w^{n}(\ulinek,\ulinem,\ulineb) \in  T_{\eta}(p_{W})\}},\mbox{ where }p_{W}(w)=\tr(\rho\mu_{w}),
\nonumber
\end{eqnarray}
which is the number of $p_{W}$-typical elements in the coset corresponding to the index $(\ulinek,\ulinem)$. We define
\begin{eqnarray}
\label{Eqn:DeefnOfbStar}
 \ulineb^{*}_{\ulinek,\ulinem} \define 
 \begin{cases} 
 \ulineb & \mbox{if }\mathrm{l}(\ulinek,\ulinem)=1\mbox{ and }w^{n}(\ulinek,\ulinem,\ulineb) \mbox{ is the unique element that is }p_{W}-\mbox{typical},\\
 \underline{0}&\mbox{if }\mathrm{l}(\ulinek,\ulinem)\neq 1.
 \nonumber
 \end{cases}
\end{eqnarray}
In other words, $\ulineb^{*}_{\ulinek,\ulinem}$ is the index of the unique element that is $p_{W}$-typical, if exactly one element is $p_{W}$-typical. Otherwise, $b^{*}_{\ulinek,\ulinem}$ is assigned $\underline{0}$. Next, we set
\begin{eqnarray}
\label{Eqn:PTPDecoderPOVMAlg}
\Delta_{\mathrm{ptp}}\define \Delta_{\mathrm{ptp},\CalY^{n}}\define \{ \Delta_{y^{n}} \define \sum_{(\ulk,\ulm) \in [\CommK]\times[\InfM]}\ketbra{\ulk~ \ulm}{\ulk~ \ulm}p_{Y|W}^{n}(y^{n}|c(\ulk,\ulm,\ulineb^{*}_{\ulk,\ulm})):y^{n}\in \mathcal{Y}^{n}\} \in \ScrM(\CalH_{K}\otimes \CalH_{M},\CalY^{n})
\end{eqnarray}
as the post-processing POVM employed by the point-to-point decoder. It is straightforward to verify $\Delta_{\mathrm{ptp}} \in \mathscr{M}(\HlbSp_{\CommK}\otimes \HlbSp_{\InfM},\mathcal{Y}^{n})$ is a POVM. From our implied notation, we have $K=q^{\kappa}, M=q^{r}$ and $B\define q^{\beta}$.
\med
\textit{Key steps and proxy operator.}
The proof follows the same proxy-state strategy as the unstructured likelihood-POVM analysis, with two changes. First, the proxy operator below need not have unit trace. Second, we use a three-term triangle inequality: the additional middle term isolates the error caused by the hidden $\ulineb$-index, which is not communicated to the decoder. For $a\in[\CommK]$, let\vspace{-0.05in}
\begin{eqnarray}
 \label{Eqn:TheAuxillaryState1AlgPOVM}
 T_{a}\define \frac{S_{a}}{MB} \mbox{ have SCD }T_{a} = \sum_{t=1}^{d^{n}}\nu_{ta}\ketbra{x_{ta}}{x_{ta}},\mbox{ i.e., }
 \braket{x_{ta}}{x_{\hatt a}} = \delta_{\hatt t}\mbox{ for } a \in [K]\mbox{ and }\sigma_{K} \define \frac{1}{\CommK}\sum_{a \in [\CommK]}T_{a}\otimes\ketbra{a}{a},\\
  \label{Eqn:DefnOfAlphaAlgPOVM}
\alpha \define \SimulationCPTP(\ketbra{\boldphi_{\sigma_{K}}}{\boldphi_{\sigma_{K}}} ) \mbox{ where }\ket{\boldphi_{\sigma_{K}}}\define \sum_{t =1}^{d^{n}}\sum_{a\in [\CommK]}\frac{1}{K}\sqrt{\nu_{ta}}\ket{\olinex_{ta}~a}\otimes \ket {x_{ta}~a}\mbox{ is the canonical purification of }\sigma_{K}.
 \end{eqnarray}
Next we define the CPTP maps that yield the intermediate operator $\alphaint$. We define the full-output POVM $\Theta_{\mathrm{ptp}}  \define \{ \tilde{\theta}_{\ulinek,\ulinem,\ulineb}  \define \theta_{\ulinek,\ulinem,\ulineb} \otimes \ketbra{\ulinek} \in \CalP(\CalH^{\otimes n} \otimes \CalH_{\CalW^{\kappa}}) : (\ulinek,\ulinem,\ulineb) \in \CalW^{\kappa}\times \CalW^{r} \times \CalW^{\beta}, \theta_{\ulinek,0}\}$ with $\theta_{\ulk,\ulm,\ulb}$ as defined in \eqref{Eqn:PTPCosetLikelihoodPOVM}. We define
\begin{eqnarray}
\label{Eqn:PTPFullDecoderPOVMAlg}
\olDelta_{\mathrm{ptp}}\define \olDelta_{\mathrm{ptp},\CalY^{n}}\define \{ \olDelta_{y^{n}}\define \! \!\!\!\!\!\sum_{(\ulk,\ulm,\ulb) \in \CalW^{\kappa}\times\CalW^{r}\times\CalW^{\beta}}\! \!\!\!\!\!\! \!\!\!\!\!\ketbra{\ulk~ \ulm~\ulb}{\ulk~ \ulm~\ulb}p_{Y|W}^{n}(y^{n}|c(\ulk,\ulm,\ulb)):y^{n}\in \mathcal{Y}^{n}\} \in \ScrM(\CalH_{K}\otimes \CalH_{M}\otimes \CalH_{B},\CalY^{n}).
\end{eqnarray}
Let
\begin{eqnarray}
 \label{Eqn:IntermediateState}
\alphaint \define \InterCPTP(\ketbra{\boldphi_{\sigma_{K}}}{\boldphi_{\sigma_{K}}} ),~\mbox{ where }
 \InterCPTP\define (i_{X}^{n}\!\otimes\!\tr_{K}\!\otimes i_{\CalY^{n}})\circ(i_{X}^{ n}\!\otimes\! i_{\mathcal{H}_{K}}\!\otimes\! \mathscr{E}^{\olDelta_{\mathrm{ptp}}}\! )\!\circ\! (i_{X}^{ n}\!\otimes\! i_{\mathcal{H}_{K}}\!\otimes\!  \mathscr{E}^{\Theta_{\mathrm{ptp}}})\!\!\!
\end{eqnarray}
and as defined before $\alphasim \define  \SimulationCPTP(\ketbra{\phi_{\rho_{\ulineK}^{\otimes n}}\!\!}{\!\phi_{\rho_{\ulineK}^{\otimes n}}\!} )$ where $\SimulationCPTP(\cdot)$ is as defined in \eqref{Eqn:AlphaSimDefnAgain}. Our proof will derive upper bounds on each of the three terms in the right hand side of 
\begin{eqnarray}
||\alpha_{\mbox{\tiny O}}-\alpha_{\mbox{\tiny S}}||_{1}&\leq& ||\alpha_{\mbox{\tiny O}}-\alphaint||_{1}+||\alphaint-\alpha||_{1}+||\alpha-\alpha_{\mbox{\tiny S}}||_{1} \nonumber\\
\label{Eqn:AlgCodesKeyTriangularIneq}
&\leq& ||\alpha_{\mbox{\tiny O}}-\alphaint||_{1}+||\alphaint-\alpha||_{1}+||\ketbra{\boldphi_{\sigma_{K}}}{\boldphi_{\sigma_{K}}} -\ketbra{\boldphi_{\rho_{K}^{\otimes n}}}{\boldphi_{\rho_{K}^{\otimes n}}}||_{1}.
\end{eqnarray}
Let us now characterize $\alphaorig$, $\alphaint$, $\alpha$ and $\alphasim$. This is the standard computation of the action of a likelihood POVM on the canonical purification of its proxy operator, applied here to $\theta_{\mathrm{ptp}}$, $\Theta_{\mathrm{ptp}}$ and the original POVM $\lambda_{\CalY}$. It gives
\begin{eqnarray}
 \label{Eqn:ProofStructLikelihoodCharOfStates1}
 \alphaorig &=&\sum_{y^{n}}\frac{\sqrt{\omega}\lambda_{y^{n}}\sqrt{\omega}}{\tr(\omega\lambda_{y^{n}})} \otimes \ketbra{y^{n}},~~ \alphaint = \sum_{\ulinek,\ulinem,\ulineb}\sum_{y^{n}}\frac{\upsilon(\ulinek,\ulinem,\ulineb)p_{Y|W}^{n}(y^{n}|w^{n}(\ulinek,\ulinem,\ulineb))}{KMB\tr(\omega\mu_{\ulinek,\ulinem,\ulineb})}{\sqrt{\omega}\mu_{\ulinek,\ulinem,\ulineb}\sqrt{\omega}}\otimes \ketbra{y^{n}}\nonumber\\
 \label{Eqn:ProofStructLikelihoodCharOfStates2}
 \alpha &=& \sum_{\ulinek,\ulinem,\ulineb}\sum_{y^{n}}\frac{\upsilon(\ulinek,\ulinem,\ulineb)p_{Y|W}^{n}(y^{n}|w^{n}(\ulinek,\ulinem,\ulinebstarkm))}{KMB\tr(\omega\mu_{\ulinek,\ulinem,\ulineb})}{\sqrt{\omega}\mu_{\ulinek,\ulinem,\ulineb}\sqrt{\omega}}\otimes \ketbra{y^{n}}. \nonumber
\end{eqnarray}
We are now set to analyze each of the terms $\norm{\alphaorig-\alphaint}_{1}$, $\norm{\alphaint-\alpha}_{1}$ and $\norm{\ketbra{\boldphi_{\sigma_{K}}}{\boldphi_{\sigma_{K}}} -\ketbra{\boldphi_{\rho_{K}^{\otimes n}}}{\boldphi_{\rho_{K}^{\otimes n}}}}_{1}$.
\med
\textit{Upper Bound on $\norm{\alphaorig-\alphaint}_{1}$} : We now employ random coding. Our generator matrices $G_{c},G_{m},G_{b}$ and the random dither vector $D^{n}$ are mutually independent and uniformly distributed. The codewords $C(\ulinek,\ulinem,\ulineb): (\ulinek,\ulinem,\ulineb) \in \CalW^{\kappa}\times \CalW^{r}\times \CalW^{\beta}$ are uniformly distributed over $\fieldq^{n}$ and pairwise independent. Suppose $\mathbf{P}$ denotes this distribution of the random code, then $\Expectation_{\mathbf{P}}\left\{ \norm{\alphaorig-\alphaint}_{1}\right\}$ is an instance of the generalized quantum covering lemma stated and proved in Sec.~\ref{Sec:QCLWithMeasureChange}. Applying Lemma~\ref{Lem:QuantumCoveringLemma} with the measure-change factor $q_X^n/p_X^n$, we have
\begin{eqnarray}
 \label{Eqn:PTPStructuredFirstTermAnalysis1}
 \Expectation_{\mathbf{P}}\left\{\norm{\alphaorig-\alphaint}_{1} \right\} \leq \exp\left\{  -n \left( \frac{(\kappa +\beta+r)\log q}{n} - I(XY;W)_{\sigma}-\log q+H(p_{W})-5\eta \right)\right\}
\end{eqnarray}
where the mutual information quantity is computed with respect to quantum state $\sigma_{XWY} \define (i_{X}\otimes \overline{\mathscr{E}}^{{Y|W}}_{p})\circ(i_{X}\otimes \mathscr{E}^{\mu})(\ketbra{\phi_{\rho}}{\phi_{\rho}})$ and $(p_{W}(w)=\tr(\mu_{w}\rho):w \in \CalW)$.

\med\textit{Upper Bound on $\norm{\alphaint-\alpha}_{1}$} : The bound on $\Expectation_{\mathbf{P}} \{\norm{\alphaint-\alpha}_{1} \}$ is obtained via the binning analysis. We provide the steps below. Note that
\begin{eqnarray}
\label{Eqn:BinningTerm1}
 \norm{\alphaint-\alpha}_{1} = \norm{ \sum_{\ulinek,\ulinem,\ulineb}\sum_{y^{n}}\frac{\upsilon(\ulinek,\ulinem,\ulineb)}{KMB\tr(\omega\mu_{\ulinek,\ulinem,\ulineb})}{\sqrt{\omega}\mu_{\ulinek,\ulinem,\ulineb}\sqrt{\omega}}\otimes \ketbra{y^{n}} \left( p_{Y|W}^{n}(y^{n}|w^{n}(\ulinek,\ulinem,\ulinebstarkm)) - p_{Y|W}^{n}(y^{n}|w^{n}(\ulinek,\ulinem,\ulineb))\right)}_{1}
 \nonumber\\
 \label{Eqn:BinningTerm2}
 \leq \norm{ \sum_{\ulinek,\ulinem,\ulineb}\sum_{y^{n}}\frac{\upsilon(\ulinek,\ulinem,\ulineb)\mathds{1}_{\{ w^{n}(\ulinek,\ulinem,\ulineb) \in T_{\eta}^{n}(p_{W})\}}}{KMB\tr(\omega\mu_{\ulinek,\ulinem,\ulineb})}{\sqrt{\omega}\mu_{\ulinek,\ulinem,\ulineb}\sqrt{\omega}}\otimes \ketbra{y^{n}} \left( p_{Y|W}^{n}(y^{n}|w^{n}(\ulinek,\ulinem,\ulinebstarkm)) - p_{Y|W}^{n}(y^{n}|w^{n}(\ulinek,\ulinem,\ulineb))\right)}_{1}
 \nonumber\\
 \label{Eqn:BinningTerm3}
 +\norm{ \sum_{\ulinek,\ulinem,\ulineb}\sum_{y^{n}}\frac{\upsilon(\ulinek,\ulinem,\ulineb)\mathds{1}_{\{ w^{n}(\ulinek,\ulinem,\ulineb) \notin T_{\eta}^{n}(p_{W})\}}}{KMB\tr(\omega\mu_{\ulinek,\ulinem,\ulineb})}{\sqrt{\omega}\mu_{\ulinek,\ulinem,\ulineb}\sqrt{\omega}}\otimes \ketbra{y^{n}} \left( p_{Y|W}^{n}(y^{n}|w^{n}(\ulinek,\ulinem,\ulinebstarkm)) - p_{Y|W}^{n}(y^{n}|w^{n}(\ulinek,\ulinem,\ulineb))\right)}_{1}
 \nonumber\\
 \label{Eqn:BinningTerm4}
 \leq \norm{ \sum_{\ulinek,\ulinem,\ulineb}\sum_{y^{n}}\frac{\upsilon(\ulinek,\ulinem,\ulineb)\mathds{1}_{\{ w^{n}(\ulinek,\ulinem,\ulineb) \in T_{\eta}^{n}(p_{W})\}}}{KMB\tr(\omega\mu_{\ulinek,\ulinem,\ulineb})}{\sqrt{\omega}\mu_{\ulinek,\ulinem,\ulineb}\sqrt{\omega}}\otimes \ketbra{y^{n}} \left( p_{Y|W}^{n}(y^{n}|w^{n}(\ulinek,\ulinem,\ulinebstarkm)) - p_{Y|W}^{n}(y^{n}|w^{n}(\ulinek,\ulinem,\ulineb))\right)}_{1}
 \nonumber\\
 \label{Eqn:BinningTerm5}
 + \sum_{\ulinek,\ulinem,\ulineb}\sum_{y^{n}}\frac{\upsilon(\ulinek,\ulinem,\ulineb)\mathds{1}_{\{ w^{n}(\ulinek,\ulinem,\ulineb) \notin T_{\eta}^{n}(p_{W})\}}}{KMB\tr(\omega\mu_{\ulinek,\ulinem,\ulineb})}\tr({\omega\mu_{\ulinek,\ulinem,\ulineb}})\left[p_{Y|W}^{n}(y^{n}|w^{n}(\ulinek,\ulinem,\ulinebstarkm))+p_{Y|W}^{n}(y^{n}|w^{n}(\ulinek,\ulinem,\ulineb)) \right] \leq \ttt_{1}+\ttt_{2}\mbox{ where}
 \nonumber
 \end{eqnarray}
 \begin{eqnarray}
  \label{Eqn:BinningTerm6}
 \ttt_{1}\define \norm{ \sum_{\ulinek,\ulinem,\ulineb}\sum_{y^{n}}\frac{\upsilon(\ulinek,\ulinem,\ulineb)\mathds{1}_{\{ w^{n}(\ulinek,\ulinem,\ulineb) \in T_{\eta}^{n}(p_{W})\}}}{KMB\tr(\omega\mu_{\ulinek,\ulinem,\ulineb})}{\sqrt{\omega}\mu_{\ulinek,\ulinem,\ulineb}\sqrt{\omega}}\otimes \ketbra{y^{n}} \left(\begin{array}{c} p_{Y|W}^{n}(y^{n}|w^{n}(\ulinek,\ulinem,\ulinebstarkm)) ~~~~\\~~~~- p_{Y|W}^{n}(y^{n}|w^{n}(\ulinek,\ulinem,\ulineb))\end{array}\right)}_{1}
 \nonumber\\
 \label{Eqn:BinningTerm7}
 \ttt_{2}\define \sum_{\ulinek,\ulinem,\ulineb}\frac{2\upsilon(\ulinek,\ulinem,\ulineb)\mathds{1}_{\{ w^{n}(\ulinek,\ulinem,\ulineb) \notin T_{\eta}^{n}(p_{W})\}}}{KMB\tr(\omega\mu_{\ulinek,\ulinem,\ulineb})}\tr({\omega\mu_{\ulinek,\ulinem,\ulineb}})
 \end{eqnarray}
We now analyze $\ttt_{1}$. Let us define $J(\ulinek,\ulinem)=\mathds{1}_{\{\mathrm{l}(\ulinek,\ulinem)=1 \}}+\mathds{1}_{\{\mathrm{l}(\ulinek,\ulinem) \neq  1, \ulinebstarkm=\uline0 \}}$. By the encoding rule, we have $J(\ulinek,\ulinem)=1$ for all $\ulinek,\ulinem$. With this, we have
\begin{eqnarray}
 \label{Eqn:BinningTerm8}
 \ttt_{1}= \norm{ \sum_{\ulinek,\ulinem,\ulineb}\sum_{y^{n}}\frac{J\upsilon(\ulinek,\ulinem,\ulineb)\mathds{1}_{\{ w^{n}(\ulinek,\ulinem,\ulineb) \in T_{\eta}^{n}(p_{W})\}}}{KMB\tr(\omega\mu_{\ulinek,\ulinem,\ulineb})}{\sqrt{\omega}\mu_{\ulinek,\ulinem,\ulineb}\sqrt{\omega}}\otimes \ketbra{y^{n}} \left( p_{Y|W}^{n}(y^{n}|w^{n}(\ulinek,\ulinem,\ulinebstarkm)) - p_{Y|W}^{n}(y^{n}|w^{n}(\ulinek,\ulinem,\ulineb))\right)}_{1}.
 \nonumber
\end{eqnarray}
We now claim
\begin{eqnarray}
  J\mathds{1}_{\{ w^{n}(\ulinek,\ulinem,\ulineb) \in T_{\eta}^{n}(p_{W})\}}\left( p_{Y|W}^{n}(y^{n}|w^{n}(\ulinek,\ulinem,\ulinebstarkm)) - p_{Y|W}^{n}(y^{n}|w^{n}(\ulinek,\ulinem,\ulineb))\right)
~~~~~~~~~~~~\nonumber\\ 
\label{Eqn:BinningTerm9}
~~~~~~~~~~~~=  \left( p_{Y|W}^{n}(y^{n}|w^{n}(\ulinek,\ulinem,\uline0)) - p_{Y|W}^{n}(y^{n}|w^{n}(\ulinek,\ulinem,\ulineb))\right)\mathds{1}_{\{\mathrm{l}(\ulinek,\ulinem)>1\}}\mathds{1}_{\{ w^{n}(\ulinek,\ulinem,\ulineb) \in T_{\eta}^{n}(p_{W})\}}
\end{eqnarray}
An argument for \eqref{Eqn:BinningTerm9} follows. We consider the different cases for $\mathrm{l}(\ulinek,\ulinem)$. Suppose $\mathrm{l}(\ulinek,\ulinem)=0$, then clearly both LHS and RHS are $0$. Indeed, when $\mathrm{l}(\ulinek,\ulinem)=0$, we have $\mathds{1}_{\{ w^{n}(\ulinek,\ulinem,\ulineb) \in T_{\eta}^{n}(p_{W})\}}=0$ and $\mathds{1}_{\{\mathrm{l}(\ulinek,\ulinem)>1\}}=0$. Suppose $\mathrm{l}(\ulinek,\ulinem)=1$, then by the encoding rule, we have $\ulinebstarkm=b$ and both RHS and LHS are zero. Suppose, $\mathrm{l}(\ulinek,\ulinem)>1$, then by the encoding rule, $\ulinebstarkm = \uline0$ and this implies RHS and LHS of \eqref{Eqn:BinningTerm9} are identical. We now use this chain of inequalities in analyzing $\ttt_{1}$. We have
\begin{eqnarray}
 \label{Eqn:BinningTerm11}
 \ttt_{1} &\leq& \sum_{\ulinek,\ulinem,\ulineb}\sum_{y^{n}}\frac{\upsilon(\ulinek,\ulinem,\ulineb)\mathds{1}_{\{ \mathrm{l}(\ulinek,\ulinem)>1,w^{n}(\ulinek,\ulinem,\ulineb) \in T_{\eta}^{n}(p_{W})\}}}{KMB\tr(\omega\mu_{\ulinek,\ulinem,\ulineb})}\tr({\omega\mu_{\ulinek,\ulinem,\ulineb}})\left[p_{Y|W}^{n}(y^{n}|w^{n}(\ulinek,\ulinem,\uline0))+p_{Y|W}^{n}(y^{n}|w^{n}(\ulinek,\ulinem,\ulineb)) \right]\nonumber\\
 \label{Eqn:BinningTerm12}
 &=& \sum_{\ulinek,\ulinem,\ulineb}\frac{2\upsilon(\ulinek,\ulinem,\ulineb)\mathds{1}_{\{ \mathrm{l}(\ulinek,\ulinem)>1,w^{n}(\ulinek,\ulinem,\ulineb) \in T_{\eta}^{n}(p_{W})\}}}{KMB\tr(\omega\mu_{\ulinek,\ulinem,\ulineb})}\tr({\omega\mu_{\ulinek,\ulinem,\ulineb}}).
  \end{eqnarray}
  Collating the bounds in \eqref{Eqn:BinningTerm7} and \eqref{Eqn:BinningTerm12}, we have
  \begin{eqnarray}
   \label{Eqn:BinningTerm13}
   \norm{\alphaint-\alpha}_{1} \leq \sum_{\ulinek,\ulinem,\ulineb}\left[\frac{2\upsilon(\ulinek,\ulinem,\ulineb)\mathds{1}_{\{\ \mathrm{l}(\ulinek,\ulinem)>1,w^{n}(\ulinek,\ulinem,\ulineb) \in T_{\eta}^{n}(p_{W})\}}}{KMB}+\frac{2\upsilon(\ulinek,\ulinem,\ulineb)\mathds{1}_{\{ w^{n}(\ulinek,\ulinem,\ulineb) \notin T_{\eta}^{n}(p_{W})\}}}{KMB}\right].
  \end{eqnarray}
We are now set to compute upper bounds on $\Expectation\{\ttT_{1}\}+\Expectation\{\ttT_{2}\}$, where $\ttT_{1}$ and $\ttT_{2}$ are the corresponding random quantities corresponding to a random code wherein $G_{c},G_{m},G_{b}$ and the random dither vector $D^{n}$ are mutually independent and uniformly distributed. Observe that the above sequence of steps have transformed the analysis to a classical information theory error event analysis. The presence of the measure transformation factor $\upsilon(\ulinek,\ulinem,\ulineb)$ implies that the above analysis is identical to a conventional binning analysis that is standard in information theory. We therefore refer the reader to \cite[Proof of Theorem~3.7]{BkNIT_PraPadShi} where a similar analysis is carried out for the event $E_{3}\cap (E_{2}')^{c}\cap (E_{1}')^{c}\cap E_{0}^{c}$. Using these steps, equivalently the usual typicality-indicator selection underlying Proposition~\ref{Prop:ClassicalJointTypicalityIndicator}, it can be verified that
\begin{eqnarray}
 \label{Eqn:BinningTermExpectation}
\Expectation\{\ttT_{1}\}+\Expectation\{\ttT_{2}\} \leq \exp\left\{-n \left(\log q-H(p_{W})-5\eta - \frac{\beta\log q}{n}\right) \right\}+\exp\left\{-5n\eta \right\}.
 \nonumber
\end{eqnarray}
\med\textit{Upper bound on $\norm{\ketbra{\boldphi_{\sigma_{K}}}{\boldphi_{\sigma_{K}}} -\ketbra{\boldphi_{\rho_{K}^{\otimes n}}}{\boldphi_{\rho_{K}^{\otimes n}}}}_{1}$} : While this is the same proxy-purification step as in Lemma~\ref{lem:canonical-purif-stability}, the fact that $\braket{\boldphi_{\sigma_{K}}}$ is not necessarily $1$ requires a refined analysis. We provide the same below. We adopt the following notational simplifications in the analysis of $\norm{\ketbra{\boldphi_{\sigma_{K}}}{\boldphi_{\sigma_{K}}} -\ketbra{\boldphi_{\rho_{K}^{\otimes n}}}{\boldphi_{\rho_{K}^{\otimes n}}}}_{1}$.

Henceforth, we let $\varphi_{P} \define \ketbra{\boldphi_{\sigma_{K}}}$ and $\varphi_{\rho_{K}} \define \ketbra{\boldphi_{\rho_{K}^{\otimes n}}}$. Our approach is to relate $\norm{\varphi_{P}-\varphi_{\rho_{K}}}_{1}$ to $\norm{\sigma_{K}-\rho_{K}^{\otimes n}}_{1}$ and recognize the latter to be another instance of the quantum soft covering lemma (Lemma \ref{Lem:QuantumCoveringLemma}). However, $\varphi_{\rho_{K}}, \varphi_{P}$ are purifications of $\rho_{K}^{\otimes n}$ and $K^{-1}\sum_{k}T_{k}\otimes\ketbra{k}$ respectively. How do we transfer an upper bound the trace distance between the states to an upper bound on the trace distance between their purifications. We do this through the relationship between trace distance and fidelity and the use of a few operator inequalities.

Let $a=1-\braket{\phi_{P}}$, $b = 4\braket{\phi_{P}}-4$ and $c = 4-\braket{\phi_{P}}{\phi_{\rho_{K}}}\braket{\phi_{\rho_{K}}}{\phi_{P}}$. Since $\braket{\phi_{\rho_{K}}}=1$, we have $b+c = 4\braket{\phi_{P}}\braket{\phi_{\rho_{K}}} -\braket{\phi_{P}}{\phi_{\rho_{K}}}\braket{\phi_{\rho_{K}}}{\phi_{P}} \geq 0$ from the Cauchy-Bunyakowski-Schwartz inequality. In conjunction with Lemmas~\ref{Prop:DistBetweenNonunitStates} and~\ref{Prop:BoundOnSqrtxPlusy}, we have
\begin{eqnarray}
 \label{Eqn:T2AnalysisRankOneBound}
 T_{2} = \norm{ \varphi_{\rho_{K}} - \varphi_{P}}_{1} \leq \sqrt{a^{2}+b+c} \leq a+\sqrt{b+c}\nonumber
\end{eqnarray}
We begin analyzing $c$. Recalling $M,K$, we have
\begin{eqnarray}
\label{Eqn:T2Analysis-1}
 c\!\!&=&\!\! 4-\braket{\phi_{P}}{\phi_{\rho_{K}}}\braket{\phi_{\rho_{K}}}{\phi_{P}} 
 \\
\label{Eqn:T2Analysis-2}
 \!\!&=&\!\! 4\left( 1- \left[ \frac{1}{K}\sum_{k}\tr(\sqrt{T_{k}}\sqrt{\omega} ) \right]^{2}\right)
 \\
\label{Eqn:T2Analysis-3}
 \!\!&\leq&\!\! 8\left( 1- \left[ \frac{1}{K}\sum_{k}\tr(\sqrt{T_{k}}\sqrt{\omega} ) \right]\right) = \frac{8}{K}\sum_{k} \zeta_{k}
\end{eqnarray}
where (i) \eqref{Eqn:T2Analysis-2} follows from substitution, (ii) the inequality in \eqref{Eqn:T2Analysis-3} follows from Lemma~\ref{Prop:BoundOn1Minusxsquared}, and the equality in \eqref{Eqn:T2Analysis-3} follows from defining $\zeta_{k} \define 1-\tr(\sqrt{T_{k}}\sqrt{\omega}) $. We have
\begin{eqnarray}
\label{Eqn:T2Analysis-4}
 \zeta_{k}\!\!&\!\!=\!\!&\!   1-\tr(\sqrt{T_{k}}\sqrt{\omega})= \tr(\rho)-\tr(\sqrt{T_{k}}\sqrt{\omega})
 \\
\label{Eqn:T2Analysis-5}
 \!\!&=&\!\!  \tr\left\{ \sqrt{\omega}\left[\sqrt{\omega}-\sqrt{T_{k}}\right]\right\}
  \leq \norm{ \sqrt{\omega}\left[\sqrt{\omega}-\sqrt{T_{k}}\right]}_{1}
  \\
\label{Eqn:T2Analysis-6}
 \!\!&\leq &\!\!  \norm{ \sqrt{\omega}}_{2}\norm{\sqrt{\omega}-\sqrt{T_{k}}}_{2} = \norm{\sqrt{\omega}-\sqrt{T_{k}}}_{2}
  \\
\label{Eqn:T2Analysis-7}
 \!\!&\leq& \norm{\sqrt{\left| \omega- T_{k} \right|}}_{2} = \sqrt{\norm{\left| \omega- T_{k} \right|}_{1}}
\end{eqnarray}
where \eqref{Eqn:T2Analysis-4} follows from $\tr(\rho)=1$, \eqref{Eqn:T2Analysis-5} follows from linearity of trace and the fact that $\tr(A)\leq \norm{A}_{1}$ for any operator $A$, the first inequality in \eqref{Eqn:T2Analysis-6} follows from \cite[Corollary IV.2.6]{BkBhatia_1997} - a form of H\"older's inequality, the equality in \eqref{Eqn:T2Analysis-6} follows from $\norm{\sqrt{\omega}}_{2}=\sqrt{\tr(\rho)}=1$, the first inequality in \eqref{Eqn:T2Analysis-6} follows from \cite[Theorem X.1.3]{BkBhatia_1997} and the equality in \eqref{Eqn:T2Analysis-6} follows from the definition of $|\cdot|$ for an operator and the definition of $\norm{\cdot}_{1}$. Substituting this upper bound on $\zeta_{k}$ in \eqref{Eqn:T2Analysis-3}, we have
\begin{eqnarray}
 \label{Eqn:T2Analysis-8}
 c \leq \frac{8}{K}\sum_{k}\sqrt{\norm{\left| \omega- T_{k} \right|}_{1}}.
  \end{eqnarray}
Having completed our analysis of $c$, we now turn to $a,b$. Recall $a=1-\braket{\phi_{P}}$, $b = 4\braket{\phi_{P}}-4$ where $\ket{\phi_{P}}$ was defined. Since $\ket{\phi_{P}}$ is a purification of $2^{-C_{1}-C_{2}}\sum_{k}T_{k}\otimes\ketbra{k}$, we have $a=1-2^{-C_{1}-C_{2}}\sum_{k}\tr(T_{k})$. If we study $\Expectation\{a\}$, we note that we are computing the $\Expectation\{\tr(T_{k})\}$. Since each entry of the $X-$table is picked IID $p_{X}$ and each entry of the $Y-$table is picked $p_{Y}$, it is a straightforward calculation to verify that $\Expectation\{\tr(T_{k})\}=1$. This implies $\Expectation\{a\}=\Expectation\{ b\}=0$. We are thus ready to evaluate the average value of $T_{2}$. Applying Jensen's inequality to the concave $\sqrt-$function, we have
\begin{eqnarray}
 \Expectation\{T_{2}\} \leq \Expectation\{a\}+\sqrt{\Expectation\{b\}+\frac{8}{K}\sum_{k}\sqrt{\Expectation{\norm{\left| \omega- T_{k} \right|}_{1}}}}
\end{eqnarray}
$\Expectation\{a\} = \Expectation\{b\} = 1-\Expectation\{\tr(T_{k})\}$ which can be easily verified to be $0$. We have thus reduced $\norm{\ketbra{\boldphi_{\sigma_{K}}}{\boldphi_{\sigma_{K}}} -\ketbra{\boldphi_{\rho_{K}^{\otimes n}}}{\boldphi_{\rho_{K}^{\otimes n}}}}_{1}$ to an instance of the quantum soft covering lemma in Lemma~\ref{Lem:QuantumCoveringLemma}.

Returning to the distributed instrument simulation problem, the preceding point-to-point construction is applied locally to the two likelihood instruments. The hidden binning indices are kept private, and the receiver replaces the point-to-point typical-representative rule by the hypothesis-testing based CQ decoder used in the proof of Theorem~\ref{Thm:DistInstSimltnOneShot}. The latter decoder recovers the algebraic component $W_{\oplus}$ from $C$ and the communicated messages. This produces the packing gain $\Gamma_{\oplus}$, whereas the local measure-change factors produce the penalties $\Delta_1$ and $\Delta_2$. Combining these packing constraints with the two local covering estimates above and then eliminating the auxiliary binning rates gives exactly \eqref{Eqn:CosetAsymptoticRegion}.
\end{proof}

\appendices

\section{\textcolor{black}{Useful Results}}
\label{AppSec:UsefulResults}

\subsection{Some Lemmas}

\begin{fulllemma}[Rank-one comparison with a non-normalized vector]
\label{Prop:DistBetweenNonunitStates}
Let $\ket{\alpha},\ket{\beta}\in\CalH$ with $\braket{\beta}=1$. Then
\begin{equation}
\label{Eqn:RankOneNonunitBound}
\bigl\|\ketbra{\alpha}-\ketbra{\beta}\bigr\|_1
\leq
\sqrt{(1-\braket{\alpha})^2+4\braket{\alpha}-4\braket{\beta}{\alpha}\braket{\alpha}{\beta}}.
\end{equation}
\end{fulllemma}

\begin{fulllemma}[Scalar square inequalities]
\label{Prop:BoundOn1Minusxsquared}
For every real number $x$, $1-x^2\leq 2(1-x)$.
\end{fulllemma}

\begin{fulllemma}[Subadditivity of the square root]
\label{Prop:BoundOnSqrtxPlusy}
For every $x,y\geq0$, $\sqrt{x+y}\leq\sqrt{x}+\sqrt{y}$.
\end{fulllemma}

\begin{fulllemma}[Closeness of canonical purifications]\label{lem:canonical-purif-stability}
Let $\rho,\sigma\in D(\mathcal H)$ and let $\ket{\varphi_\rho},\ket{\varphi_\sigma}$ denote their canonical purifications (with respect to a fixed orthonormal basis).
Then
\[
\big\|\,\ket{\varphi_\rho}\bra{\varphi_\rho}-\ket{\varphi_\sigma}\bra{\varphi_\sigma}\,\big\|_1 \;\le\; 2\sqrt{2}\;\big\|\rho-\sigma\big\|_1^{1/4}.
\]
This is Lemma~8 in Appendix~A of~\cite{2026MMTIT_Pad}.
\end{fulllemma}
\begin{fulllemma}[Gentle measurement lemma]\label{Lem:GentleMeasurement}
Let $\rho$ be a density operator on a finite-dimensional Hilbert space $\CalH$, and let
$0 \preceq \Lambda \preceq \mathbb{I}$ be a POVM element.
Set $p \define \Tr[\Lambda \rho]$. Then
\begin{equation}
\label{Eqn:GentleMeasurement}
\bigl\|\rho - \sqrt{\Lambda}\,\rho\,\sqrt{\Lambda}\bigr\|_1
\;\le\;
2\sqrt{1-p}.
\end{equation}
\cite[Lemma 9.4.2]{BkWilde_2017}
\end{fulllemma}

\begin{fulllemma}[Trace distance to a pure state]
\label{Lem:TraceDistancePureState}
For any density operator $\sigma$ and any pure state $\ket{\phi}$, we have
\begin{equation}
\label{Eqn:TraceDistancePureStateBounds}
2\left(1-\sqrt{\bra{\phi}\sigma\ket{\phi}}\right)
\;\le\;
\bigl\|\sigma-\ket{\phi}\bra{\phi}\bigr\|_1
\;\le\;
2\sqrt{1-\bra{\phi}\sigma\ket{\phi}}.
\end{equation}
This is Eq.~(9.164) in \cite{BkWilde_2017}.
\end{fulllemma}

\subsection{One-shot Two-party Soft Covering in Expectation}

In~\cite{202410arXiv_Sen},
Sen proved fully smooth one-shot multipartite convex-split (soft-covering) bounds, based on a telescoping (mean-zero) decomposition
that bypasses the simultaneous smoothing bottleneck.
We will use these bounds as a key tool in the error analysis of our protocol.

\begin{fullproposition}[Smooth multipartite soft covering in expectation]
\label{Prop:SmoothMultipartiteSoftCovering}
Let $k$ be a positive integer. Let $\mathcal{X}_1,\dots,\mathcal{X}_k$ be $k$ classical alphabets.
For any subset $S\subseteq[k]$, let $\mathcal{X}_S := (\mathcal{X}_s)_{s\in S}$.
Let $p_{\mathcal{X}_{[k]}}$ be a normalized probability distribution on
$\mathcal{X}_{[k]} := \mathcal{X}_1\times\cdots\times \mathcal{X}_k$,
and let $p_{\mathcal{X}_S}$ denote its marginal distribution on $\mathcal{X}_S$.
Let $q_{\mathcal{X}_1},\dots,q_{\mathcal{X}_k}$ be normalized probability distributions on the respective alphabets.
For each $(x_1,\dots,x_k)\in\mathcal{X}_{[k]}$, let $\rho^M_{x_1,\dots,x_k}$ be a subnormalized density matrix on $M$.
Define the classical-quantum control state
\[
\rho^{\mathcal{X}_{[k]}M}
:= \sum_{(x_1,\dots,x_k)\in\mathcal{X}_{[k]}}
p_{\mathcal{X}_{[k]}}(x_1,\dots,x_k)\,
\ket{x_1,\dots,x_k}\bra{x_1,\dots,x_k}^{\mathcal{X}_{[k]}}
\otimes \rho^M_{x_1,\dots,x_k}.
\]
Suppose $\supp(p_{\mathcal{X}_i})\subseteq \supp(q_{\mathcal{X}_i})$ for all $i\in[k]$.
For any subset $S\subseteq[k]$, let $q_{\mathcal{X}_S} := \times_{s\in S} q_{\mathcal{X}_s}$.
Let $A_1,\dots,A_k$ be positive integers.
For each $i\in[k]$, let $x_i^{(A_i)} := (x_i(1),\dots,x_i(A_i))$ denote an $A_i$-tuple of elements from $\mathcal{X}_i$.
Denote the $A_i$-fold product alphabet $\mathcal{X}_i^{A_i} := \mathcal{X}_i^{\times A_i}$,
and the product probability distribution $q_{\mathcal{X}_i^{A_i}} := (q_{\mathcal{X}_i})^{\times A_i}$.
For any collection of tuples $x_i^{(A_i)}\in \mathcal{X}_i^{A_i}$, $i\in[k]$, define the sample-average covering state
\[
\sigma^M_{x_1^{(A_1)},\dots,x_k^{(A_k)}}
:= (A_1\cdots A_k)^{-1}
\sum_{a_1=1}^{A_1}\cdots\sum_{a_k=1}^{A_k}
\frac{p_{\mathcal{X}_{[k]}}\bigl(x_1(a_1),\dots,x_k(a_k)\bigr)}
{q_{\mathcal{X}_1}\bigl(x_1(a_1)\bigr)\cdots q_{\mathcal{X}_k}\bigl(x_k(a_k)\bigr)}
\;\rho^M_{x_1(a_1),\dots,x_k(a_k)}.
\]
Suppose that for each non-empty subset $\varnothing\neq S\subseteq[k]$,
\[
\sum_{s\in S}\log A_s
>
D_2^{\varepsilon}\Bigl(\rho^{\mathcal{X}_S M}\,\big\|\, q_{\mathcal{X}_S}\otimes \rho^M\Bigr)
+\log\varepsilon^{-2}.
\]
Then,
\[
\mathbb{E}_{x_1^{(A_1)},\dots,x_k^{(A_k)}}
\Bigl[\bigl\|\sigma^M_{x_1^{(A_1)},\dots,x_k^{(A_k)}}-\rho^M\bigr\|_1\Bigr]
<
2(3k-1)\varepsilon\,(\Tr\rho),
\]
where the expectation is taken over independent choices of tuples $x_i^{(A_i)}$
from the distributions $q_{\mathcal{X}_i^{A_i}}$, $i\in[k]$.
\end{fullproposition}

\begin{fulldefinition}[Worst-case conditional $I_2$ on a high-probability set]
\label{Defn:WorstCaseConditionalI2}
Let $\underline{\CalW}\define \CalW_1\times\CalW_2$ and consider a cq-state of the form
\[
\sigma^{\underline{W}W_0A}
=
\sum_{\underline{w}\in\underline{\CalW}} p_{\underline{W}}(\underline{w})\,\ketbra{\underline{w}}_{\underline{W}}
\otimes
\sum_{w_0\in\CalW_0} p_{W_0|\underline{W}}(w_0|\underline{w})\,\ketbra{w_0}_{W_0}\otimes \rho^{w_0,\underline{w}}_A,
\]
and define, for each $\underline{w}\in\underline{\CalW}$, the conditional slice
\[
\sigma_{\underline{w}}^{W_0A}
\define
\sum_{w_0\in\CalW_0} p_{W_0|\underline{W}}(w_0|\underline{w})\,\ketbra{w_0}_{W_0}\otimes \rho_A^{w_0,\underline{w}}.
\]
For $\varepsilon_0,\varepsilon\in(0,1)$, define
\begin{equation}
\label{Eqn:DefWorstCaseConditionalI2}
I_{2,\mathrm{wc}}^{\varepsilon_0,\varepsilon}(W_0\!:\!A\,|\,\underline{W})_{\sigma}
\define
\min_{\Omega\subseteq\underline{\CalW}:\ p_{\underline{W}}(\Omega)\ge 1-\varepsilon_0}
\ \max_{\underline{w}\in\Omega}\ 
I_2^{\varepsilon}(W_0\!:\!A)_{\sigma_{\underline{w}}}.
\end{equation}
\end{fulldefinition}

\begin{fulllemma}[Conditional one-shot soft covering]
\label{Lem:ConditionalSoftCovering}
Fix $\varepsilon_0,\varepsilon\in(0,1)$ and let $\sigma^{\underline{W}W_0A}$ be as in
Definition~\ref{Defn:WorstCaseConditionalI2}.
For each $\underline{w}\in\underline{\CalW}$, generate i.i.d.\ codewords
$W_0(1),\dots,W_0(M_0)\sim p_{W_0|\underline{W}}(\cdot|\underline{w})$
and define the random mixture
\[
\widetilde{\rho}^{\,\underline{w}}_A \define \frac{1}{M_0}\sum_{m_0=1}^{M_0}\rho_A^{W_0(m_0),\underline{w}},
\qquad
\overline{\rho}^{\,\underline{w}}_A \define \sum_{w_0\in\CalW_0} p_{W_0|\underline{W}}(w_0|\underline{w})\,\rho_A^{w_0,\underline{w}}.
\]
If
\begin{equation}
\label{Eqn:CondSC_RateCond}
\log M_0
\ \ge\
I_{2,\mathrm{wc}}^{\varepsilon_0,\varepsilon}(W_0\!:\!A\,|\,\underline{W})_{\sigma}
\ +\ 2\log\frac{1}{\varepsilon^2},
\end{equation}
then
\begin{equation}
\label{Eqn:CondSC_Bound}
\mathbb{E}_{\underline{W}\sim p_{\underline{W}}}\,
\mathbb{E}_{\mathcal{C}_0|\underline{W}}\!\left[
\bigl\|\widetilde{\rho}^{\,\underline{W}}_A-\overline{\rho}^{\,\underline{W}}_A\bigr\|_1
\right]
\ \le\
2\varepsilon_0+4\varepsilon.
\end{equation}
\end{fulllemma}

\begin{proof}
Let $\Omega^\star\subseteq\underline{\CalW}$ attain the minimum in \eqref{Eqn:DefWorstCaseConditionalI2}, so that $p_{\underline{W}}(\Omega^\star)\ge 1-\varepsilon_0$ and
\[
\max_{\underline{w}\in\Omega^\star} I_2^{\varepsilon}(W_0\!:\!A)_{\sigma_{\underline{w}}}
=
I_{2,\mathrm{wc}}^{\varepsilon_0,\varepsilon}(W_0\!:\!A\,|\,\underline{W})_{\sigma}.
\]
Split the expectation in \eqref{Eqn:CondSC_Bound} as
\[
\sum_{\underline{w}\in\Omega^\star}p_{\underline{W}}(\underline{w})\,
\mathbb{E}_{\mathcal{C}_0|\underline{w}}\!\bigl[\|\widetilde{\rho}^{\,\underline{w}}_A-\overline{\rho}^{\,\underline{w}}_A\|_1\bigr]
+
\sum_{\underline{w}\notin\Omega^\star}p_{\underline{W}}(\underline{w})\,
\mathbb{E}_{\mathcal{C}_0|\underline{w}}\!\bigl[\|\widetilde{\rho}^{\,\underline{w}}_A-\overline{\rho}^{\,\underline{w}}_A\|_1\bigr].
\]
The second sum is bounded by $2\,p_{\underline{W}}((\Omega^\star)^c)\le 2\varepsilon_0$ since
$\|\widetilde{\rho}^{\,\underline{w}}_A-\overline{\rho}^{\,\underline{w}}_A\|_1\le 2$.
For $\underline{w}\in\Omega^\star$, the standard one-shot soft-covering bound (applied to the cq-state
$\sigma_{\underline{w}}^{W_0A}$ and rate condition \eqref{Eqn:CondSC_RateCond}) yields
\[
\mathbb{E}_{\mathcal{C}_0|\underline{w}}\!\bigl[\|\widetilde{\rho}^{\,\underline{w}}_A-\overline{\rho}^{\,\underline{w}}_A\|_1\bigr]
\le 4\varepsilon.
\]
Combining both contributions gives \eqref{Eqn:CondSC_Bound}.
\end{proof}

\subsection{One-shot Classical Joint Typicality}

\begin{fullproposition}[One-shot classical joint typicality]
\label{Prop:ClassicalJointTypicalityIndicator}
Let $\mathcal{X}$ be a finite set.
Let $\{p_i\}_{i=1}^t$ and $\{q_j\}_{j=1}^\ell$ be probability distributions on $\mathcal{X}$,
and let $\{\varepsilon_{ij}\}_{i\in[t],\,j\in[\ell]}\subset(0,1)$.
Then there exists a function
\[
f:\mathcal{X}\longrightarrow\{0,1\}
\]
such that
\begin{enumerate}
\item for all $i\in[t]$,
\[
\sum_{x\in\mathcal{X}} p_i(x)\,f(x)
\;\ge\;
1-\sum_{j=1}^{\ell}\varepsilon_{ij},
\]
\item for all $j\in[\ell]$,
\[
\sum_{x\in\mathcal{X}} q_j(x)\,f(x)
\;\le\;
\sum_{i=1}^{t} 2^{-D_H^{\varepsilon_{ij}}(p_i\|q_j)}.
\]
\end{enumerate}
\end{fullproposition}

\section{Compatible Operator Sliding for Theorem~\ref{Thm:DistInstSimltnOneShot}}
\label{AppSec:QSI-COS}

This appendix records the compatible-operator-sliding computation used in the proof of Theorem~\ref{Thm:DistInstSimltnOneShot}. The purpose is to justify the replacement of the likelihood encoding instrument, which contains the random inverse $S_{k_i}^{-1/2}$, by the inverse-free map $\widetilde{\CalI}_{k_i}^{(i)}$ in \eqref{Eqn:QSIAuxiliaryMaps}. The key point is that when the likelihood instrument acts on the canonical purification of $S_{k_i}$, the factor $S_{k_i}^{-1/2}$ cancels against the $\sqrt{S_{k_i}}$ already present in the purification.

\begin{fulllemma}[Single-link compatible-operator sliding]
\label{Lem:QSICOSSingleLink}
Fix $i\in[2]$ and a common-randomness index $k_i$. Let $\mathsf{V}_i$ be the isometry in \eqref{Eqn:QSIIsometricExtensionState}. Then
\begin{eqnarray}
\label{Eqn:QSICOSSingleLink}
\left(\mathrm{id}_{XCA_{3-i}}\otimes \CalE_{k_i}^{(i)}\right)
\left(
(\mathsf{V}_i\otimes I_{A_i})\,
\varphi_{S_{k_i}}\,
(\mathsf{V}_i^*\otimes I_{A_i})
\right)
=
\left(\mathrm{id}_{XCA_{3-i}}\otimes \widetilde{\CalI}_{k_i}^{(i)}\right)
\left(\varphi^{XA_1A_2C}\right).
\end{eqnarray}
\end{fulllemma}

\begin{proof}
We give the proof for a fixed $i$; the two cases are identical. By the definition of the encoding instrument in \eqref{Eqn:EncodingInstrumentki}, the left-hand side of \eqref{Eqn:QSICOSSingleLink} is
\begin{eqnarray}
\Tr_{E_i}\!\left[\sum_{m_i,b_i}
\left(\mathrm{id}_{XCA_{3-i}}\otimes \theta_{k_i}^{m_i,b_i}\right)
\left(
(\mathsf{V}_i\otimes I_{A_i})\varphi_{S_{k_i}}(\mathsf{V}_i^*\otimes I_{A_i})
\right)
\left(\mathrm{id}_{XCA_{3-i}}\otimes {\theta_{k_i}^{m_i,b_i}}^\dagger\right)
\otimes \ketbra{m_i}_{M_i}
\right]
\nonumber
\end{eqnarray}
plus the analogous term with $\theta^0_{k_i}$ and the message $0$. The latter term is zero. Indeed, if
\[
\varphi_{S_{k_i}}
=
\sum_{t,t'}
\ket{\overline{e_t}}\bra{\overline{e_{t'}}}
\otimes
\sqrt{S_{k_i}}\ket{e_t}\bra{e_{t'}}\sqrt{S_{k_i}},
\]
then \eqref{Eqn:DistThetak0GetsKilled_i} gives $\theta^0_{k_i}\sqrt{S_{k_i}}=0$.

For the nonzero terms, $\mathsf{V}_i$ acts on the purifying system while $\theta_{k_i}^{m_i,b_i}$ acts on $A_i$. Hence the compatible operators slide past each other:
\[
\left(\mathrm{id}_{XCA_{3-i}}\otimes \theta_{k_i}^{m_i,b_i}\right)
(\mathsf{V}_i\otimes I_{A_i})
=
(\mathsf{V}_i\otimes I_{B_iE_i})
\left(\mathrm{id}_{R_i}\otimes \theta_{k_i}^{m_i,b_i}\right),
\]
where $R_i$ denotes the purifying register of $\varphi_{S_{k_i}}$ and $\varphi_{\rho_{A_i}}$. Using again the canonical expansion of $\varphi_{S_{k_i}}$ and the identity in \eqref{Eqn:DistThetak0GetsKilled_i}, we get
\begin{eqnarray}
\left(\mathrm{id}_{R_i}\otimes \theta_{k_i}^{m_i,b_i}\right)
\varphi_{S_{k_i}}
\left(\mathrm{id}_{R_i}\otimes {\theta_{k_i}^{m_i,b_i}}^\dagger\right)
=
\frac{
\left(\mathrm{id}_{R_i}\otimes V_{w_i(k_i,m_i,b_i)}\right)
\varphi_{\rho_{A_i}}
\left(\mathrm{id}_{R_i}\otimes V_{w_i(k_i,m_i,b_i)}^\dagger\right)
}{
M_iB_i\,p_{W_i}(w_i(k_i,m_i,b_i))
}.
\label{Eqn:QSICOSCancellation}
\end{eqnarray}
Applying $\mathsf{V}_i$ on the purifying side and using \eqref{Eqn:QSIIsometricExtensionState} transforms $\varphi_{\rho_{A_i}}$ into $\varphi^{XA_1A_2C}$. Substituting \eqref{Eqn:QSICOSCancellation} into the expanded encoding instrument gives exactly the right-hand side of \eqref{Eqn:QSICOSSingleLink}, by the definition of $\widetilde{\CalI}_{k_i}^{(i)}$ in \eqref{Eqn:QSIAuxiliaryMaps}.
\end{proof}

Applying Lemma~\ref{Lem:QSICOSSingleLink} with $i=1$ to the state $\CalT_3$ in \eqref{Eqn:QSIT3BeforeCOS} gives the representation \eqref{Eqn:QSICOST3After}. The same calculation applied inside the proxy state $\CalT_2$, before the final CQMAC decoding and stochastic post-processing, gives \eqref{Eqn:QSICOST2After}. In both cases the decoder is outside the sliding step and is therefore unaffected by the calculation.

\begin{fulllemma}[Averaged contraction of the inverse-free map]
\label{Lem:QSIInverseFreeMapContraction}
Let $F$ be any finite-dimensional reference system and let $L$ be a Hermitian operator on $\CalH_F\otimes\CalH_{A_i}$. For every fixed $k_i$,
\begin{equation}
\label{Eqn:QSIInverseFreeMapContraction}
\Expectation_{\CalC_i}\!\left[
\left\|
(\mathrm{id}_{F}\otimes\widetilde{\CalI}_{k_i}^{(i)})(L)
\right\|_1
\right]
\leq
\norm{L}_1,
\end{equation}
where the expectation is over the random codebook of sender $i$.
\end{fulllemma}

\begin{proof}
By convexity of the trace norm and the definition of $\widetilde{\CalI}_{k_i}^{(i)}$,
\begin{eqnarray}
\Expectation_{\CalC_i}\!\left[
\left\|
(\mathrm{id}_{F}\otimes\widetilde{\CalI}_{k_i}^{(i)})(L)
\right\|_1
\right]
&\leq&
\Expectation_{C_i}\!\left[
\left\|
\frac{(\mathrm{id}_{F}\otimes V_{_{w_i(k_i,1,1)}})L(\mathrm{id}_{F}\otimes V_{_{w_i(k_i,1,1)}}^\dagger)}
{p_{W_i}(W_i)}
\right\|_1
\right]
\nonumber\\
&=&
\sum_{w_i\in\CalW_i}
\left\|
(\mathrm{id}_{F}\otimes V_{w_i})L(\mathrm{id}_{F}\otimes V_{w_i}^\dagger)
\right\|_1
\nonumber\\
&=&
\left\|
\sum_{w_i\in\CalW_i}
(\mathrm{id}_{F}\otimes V_{w_i})L(\mathrm{id}_{F}\otimes V_{w_i}^\dagger)
\otimes\ketbra{w_i}
\right\|_1
\nonumber\\
&\leq&
\norm{L}_1.
\end{eqnarray}
The last inequality follows because the map
$L\mapsto\sum_{w_i}(\mathrm{id}_{F}\otimes V_{w_i})L(\mathrm{id}_{F}\otimes V_{w_i}^\dagger)\otimes\ketbra{w_i}$
is completely positive and trace preserving on the $A_i$ system, since $\sum_{w_i}V_{w_i}^\dagger V_{w_i}=I_{A_i}$.
\end{proof}

\section{Quantum Covering Lemma}
\label{Sec:QCLWithMeasureChange}
We formulate and prove a slightly general quantum covering \cite[Chap.~17]{BkWilde_2017} that permits choosing the random density operators according to a different distribution than that is used in the averaging.

\begin{fulllemma}
\label{Lem:QuantumCoveringLemma}
 Let $\mathcal{H}$ be a finite dimensional Hilbert space of dimension $d$, $\CalX$ be a finite set and for each $x \in \CalX$, let $\rho_{x} \in \CalD(\CalH)$ be density operators. Let $p_{X}$ and $q_{X}$ be two distributions on $\CalX$ and $\sigma = \sum_{u \in \CalX}q_{X}(u)\rho_{u}$. Suppose $C=(X^{n}(m) \in \CalX^{n}:1\leq m \leq M)$ be a collection of $M$ pairwise independent and identically distributed vectors with $\mathbf{P}(X^{n}(m) = x^{n}) = \prod_{t=1}^{n}p_{X}(x_{t})$ for each $m \in [M]$. Then, there exists $\eta > 0$ such that for sufficiently large $n \in \naturals$, we have
 \begin{eqnarray}
  \label{Eqn:GeneralQCL}
  \mathbb{E}_{\mathbf{P}}\left\{ \norm{\sigma^{\otimes n} - A   }_{1} \right\} \leq \exp\left\{ -n  \left[ \frac{\log M}{n} +\sum_{x\in \CalX}q_{x}(x)S(\rho_{x})-D(q_{X}||p_{X})[1 +4\eta]-S(\rho)\right] \right\}, \nonumber\\ \mbox{ where }A \define \frac{1}{M}\sum_{m=1}^{M}\!\!A(m), ~A(m) \define \frac{q_{X}^{n}(X^{n}(m))}{p_{X}^{n}(X^{n}(m))}\rho_{X^{n}(m)}, ~~\rho_{x^{n}}=\displaystyle \bigotimes_{t=1}^{n}\rho_{x_{t}} ~\mbox{ for any $x^{n} = (x_{1},\cdots,x_{n}) \in \CalX^{n}$.}
  \nonumber
 \end{eqnarray}
\end{fulllemma}
\begin{proof}
Before we step into the outline and details of the proof, let us identify the unconditional and conditional typical projectors. Towards that end, let $\CalY = [d]$ and for $x \in \CalX$, let $\rho_{x} = \sum_{y \in \CalY}r_{Y|X}(y|x)\ketbra{e_{y|x}}$ be a spectral decomposition. Similarly, let $\sigma = \sum_{x \in \CalX}q_{X}(x)\rho_{x}$ admit a spectral decomposition $\sigma = \sum_{y \in \CalY}s_{Y}(y)\ketbra{f_{y}}$. We define the conditional typical projector with respect to the joint distribution $q_{X}r_{Y|X}$ and the (unconditional) typical projectors with respect to the distribution $s_{Y}$ as
\begin{eqnarray}
\label{Eqn:TypicalProjectors}
 \pi_{x^{n}} \define \sum_{y^{n} \in \CalY^{n}}\displaystyle \bigotimes_{t=1}^{n}\ketbra{e_{y_{t}|x_{t}}}\mathds{1}_{\{(x^{n},y^{n}) \in T_{\eta} (q_{X}r_{Y|X})\}},~~ \pi_{m} \define \pi_{X^{n}(m)}, ~~ \pi \define \sum_{y^{n} \in \CalY^{n}}\displaystyle \bigotimes_{t=1}^{n}\ketbra{f_{y_{t}}}\mathds{1}_{\{y^{n} \in T_{4\eta} (s_{Y})\}}.
 \nonumber
\end{eqnarray}

Our technique is based on Cuff's proof of classical soft covering \cite[Lemma 19]{CuffPhDThesis}. We recognize that $\overline{A} \define \Expectation_{\mathbf{P}}\{ A\} =\sigma^{\otimes n}$ and hence the quantity of interest is $\mathbb{E}_{\mathbf{P}}\left\{ \norm{\overline{A} - A   }_{1} \right\}$. We identify three terms $T_{1},T_{2}$ and $T_{3}$ given by $T_{1}= \mathbb{E}_{\mathbf{P}}\left\{ \norm{{A} - {B}   }_{1} \right\}$, $T_{2} = \mathbb{E}_{\mathbf{P}}\left\{ \norm{B - \overline{B}   }_{1} \right\}$ and $T_{3} = \mathbb{E}_{\mathbf{P}}\left\{ \norm{\overline{B} - \overline{A}   }_{1} \right\}$ where $A$ is as defined in the theorem statement, 
\begin{eqnarray}
 \label{Eqn:QCLProofPrelims1}
 B =\!\sum_{m=1}^{M}\!\!\frac{B(m)}{M}, ~B(m)  \define \frac{q_{X}^{n}(X^{n}(m))}{p_{X}^{n}(X^{n}(m))}\pi\pi_{m}\rho_{X^{n}(m)}\pi_{m}\pi,~ \overline{B} \define \Expectation_{\mathbf{P}}\{A\}\mbox{ and we recognize }\mathbb{E}_{\mathbf{P}}\left\{ \norm{\overline{A} - A   }_{1} \right\} \leq T_{1}+T_{2}+T_{3}.
 \nonumber
\end{eqnarray}
Our proof involves analyzing each of $T_{1},T_{2}$ and $T_{3}$. As we shall see $T_{1},T_{3}$ can be bounded using properties of unconditional and conditional typical projectors and $T_{2}$ is the important term yielding our rate constraints, i.e., the lower bound on $\frac{\log M}{n}$. We therefore begin analyzing $T_{2}$.

\med\textit{An Upper Bound on $T_{2}$ : } For ease of reference, we recall $T_{2} = \mathbb{E}_{\mathbf{P}}\left\{ \norm{B - \overline{B}   }_{1} \right\}$. As in the proof of \cite[Lemma 19]{CuffPhDThesis}, we leverage the concavity and monotonicity properties of the square-root function. While this is rather straightforward for real numbers in \cite[Lemma 19]{CuffPhDThesis}, we require operator monotonicity and operator concavity of the square root function. As proven in the Lowner-Heinz theorem \cite[Theorem~2.6]{201001CM_Car}, this is indeed true. The operator square root function is operator monotonic and operator concave. We therefore have
\begin{eqnarray}
\label{Eqn:GenQCLSimplifyingT2_1}
T_{2} &=& \Expectation_{\mathbf{P}}\left\{ \norm{B-\overline{B}}_{1}\right\} = \Expectation_{\mathbf{P}}\left\{ \tr(\sqrt{\left[B-\overline{B}\right]\left[B-\overline{B}\right]^{\dagger}}) \right\} =  \tr(\Expectation_{\mathbf{P}}\left\{ \sqrt{\left[B-\overline{B}\right]\left[B-\overline{B}\right]^{\dagger}}\right\}) \nonumber\\
\label{Eqn:GenQCLSimplifyingT2_2}
&\leq& \tr( \sqrt{\Expectation_{\mathbf{P}}\left\{ \left[B-\overline{B}\right]\left[B-\overline{B}\right]^{\dagger}\right\}}) =  \tr( \sqrt{\Expectation_{\mathbf{P}}\left\{ BB^{\dagger}\right\} - \Expectation_{\mathbf{P}}\left\{ B \right\} \Expectation_{\mathbf{P}}\left\{ B^{\dagger} \right\} })
\end{eqnarray}
where the last equality follows from $\Expectation_{\mathbf{P}}\left\{ B\overline{B}^{\dagger}\right\} = \Expectation_{\mathbf{P}}\left\{ \overline{B}{B}^{\dagger}\right\}  = \overline{B}~\overline{B}^{\dagger} = \Expectation_{\mathbf{P}}\left\{ B \right\} \Expectation_{\mathbf{P}}\left\{ B^{\dagger} \right\}$. To reduce clutter in notation, we henceforth abbreviate $p_{m} = p_{X}^{n}(X^{n}(m))$, $q_{m} = q_{X}^{n}(X^{n}(m))$, $\rho_{m} = \rho_{X^{n}(m)}, \zeta_{m}=\frac{p_{m}}{q_{m}}$. With this and recognizing the Hermitian nature of B, we have
\begin{eqnarray}
 \label{Eqn:GenQCLSimplifyingT2_3}
 BB^{\dagger} \!= \!\sum_{m=1}^{M}\frac{\zeta_{m}^{2}\pi\pi_{m}\rho_{m}\pi_{m}\pi\pi\pi_{m}\rho_{m}\pi_{m}\pi}{M^{2}} + \!\sum_{\substack{a,b=1\\b\neq a}}^{M}\frac{\zeta_{a}\zeta_{b}\pi\pi_{a}\rho_{a}\pi_{a}\pi\pi\pi_{b}\rho_{b}\pi_{b}\pi}{M^{2}},~ \Expectation_{\mathbf{P}}\{B \} = \!\!\!\sum_{u^n\in \CalX^{n}}\!\!\!\!q_{X}^{n}(u^{n})\pi\pi_{u^{n}}\rho_{u^{n}}\pi_{u^{n}}\pi
 \end{eqnarray}
where we have used the identically distributed nature of $X^{n}(m): m\in[ M]$ in obtaining the last equality above. The pairwise independent nature of the distribution of the collection $(X^{n}(m): 1\leq m \leq M)$ implies
\begin{eqnarray}
 \label{Eqn:GenQCLSimplifyingT2_4}
 \lefteqn{\Expectation_{\mathbf{P}}\left\{ \!\sum_{\substack{a,b=1\\b\neq a}}^{M}\frac{\zeta_{a}\zeta_{b}\pi\pi_{a}\rho_{a}\pi_{a}\pi\pi\pi_{b}\rho_{b}\pi_{b}\pi}{M^{2}} \right\} = \!\sum_{\substack{a,b=1\\b\neq a}}^{M}\frac{\Expectation_{\mathbf{P}}\left\{\zeta_{a}\pi\pi_{a}\rho_{a}\pi_{a}\pi\right\}\Expectation\left\{\zeta_{b}\pi\pi_{b}\rho_{b}\pi_{b}\pi \right\}}{M^{2}}} \nonumber\\
  &&~~~~~~=\sum_{\substack{a,b=1\\b\neq a}}^{M}\sum_{\substack{x^{n},\tilde{x}^{n}  \\\in \CalX^{n}}}\frac{q_{X}^{n}(x^{n})q_{X}^{n}(\tilde{x}^{n})}{M^{2}}\pi\pi_{x^{n}}\rho_{x^{n}}\pi_{x^{n}}\pi\pi\pi_{\tilde{x}^{n}}\rho_{\tilde{x}^{n}}\pi_{\tilde{x}^{n}}\pi
 \label{Eqn:GenQCLSimplifyingT2_6}
 =\frac{M^{2}-M}{M^{2}}\left(\sum_{\substack{x^{n} \in \\\CalX^{n}}}q_{X}^{n}(x^{n})\pi\pi_{x^{n}}\rho_{x^{n}}\pi_{x^{n}}\pi \right)^{2}
\end{eqnarray}
Substituting \eqref{Eqn:GenQCLSimplifyingT2_3} and \eqref{Eqn:GenQCLSimplifyingT2_6} in 
\begin{eqnarray}
\label{Eqn:GenQCLSimplifyingT2_7}
 \Expectation_{\mathbf{P}}\left\{ BB^{\dagger}\right\} - \Expectation_{\mathbf{P}}\left\{ B \right\} \Expectation_{\mathbf{P}}\left\{ B^{\dagger} \right\} = \Expectation_{\mathbf{P}}\left\{ \!\frac{1}{M^{2}}\!\!\!\sum_{m=1}^{M}\zeta_{m}^{2}\pi\pi_{m}\rho_{m}\pi_{m}\pi\pi\pi_{m}\rho_{m}\pi_{m}\pi\right\} - \frac{1}{M}\left(\sum_{\substack{x^{n} \in \\\CalX^{n}}}q_{X}^{n}(x^{n})\pi\pi_{x^{n}}\rho_{x^{n}}\pi_{x^{n}}\pi \right)^{2} .
\end{eqnarray}
Since $\displaystyle\sum_{{x^{n} \in \CalX^{n}}}q_{X}^{n}(x^{n})\pi\pi_{x^{n}}\rho_{x^{n}}\pi_{x^{n}}\pi$ is positive semi-definite, the second term in \eqref{Eqn:GenQCLSimplifyingT2_7} is positive-semidefinite. From operator monotonicity of square root function \cite[Theorem~2.6]{201001CM_Car}, we have
\begin{eqnarray}
\label{Eqn:GenQCLSimplifyingT2_8} 
\sqrt{\Expectation_{\mathbf{P}}\left\{ BB^{\dagger}\right\} - \Expectation_{\mathbf{P}}\left\{ B \right\} \Expectation_{\mathbf{P}}\left\{ B^{\dagger} \right\}} \leq \sqrt{\Expectation_{\mathbf{P}}\left\{ \!\frac{1}{M^{2}}\!\!\!\sum_{m=1}^{M}\zeta_{m}^{2}\pi\pi_{m}\rho_{m}\pi_{m}\pi\pi\pi_{m}\rho_{m}\pi_{m}\pi\right\}}~~~~\mbox{ and hence }\nonumber\\
\label{Eqn:GenQCLSimplifyingT2_9}
T_{2} \leq \tr( \sqrt{\Expectation_{\mathbf{P}}\left\{ BB^{\dagger}\right\} - \Expectation_{\mathbf{P}}\left\{ B \right\} \Expectation_{\mathbf{P}}\left\{ B^{\dagger} \right\} }) \leq \frac{1}{M}\tr(\sqrt{\sum_{m=1}^{M}\Expectation_{\mathbf{P}} \left\{\zeta_{m}^{2}\pi\pi_{m}\rho_{m}\pi_{m}\pi\pi\pi_{m}\rho_{m}\pi_{m}\pi\right\}}).
\end{eqnarray}
In the following, we focus on bounding the operator under the square root in (\ref{Eqn:GenQCLSimplifyingT2_9}) through a sequence of operator inequalities. Using the basic facts (i) $ABA^{\dagger} \leq ACA^{\dagger}$ whenever $B \leq C$, (ii) $\pi^{2} = \pi \leq I$, (iii) $\pi_{x^{n}}^{2} = \pi_{x^{n}} \leq I$ we have
\begin{eqnarray}
 \label{Eqn:GenQCLSimplifyingT2_10}
 &&\!\!\!\!\!\!\!\!\!\!\!\!\!\!\Expectation_{\mathbf{P}} \left\{\zeta_{m}^{2}\pi\pi_{m}\rho_{m}\pi_{m}\pi\pi\pi_{m}\rho_{m}\pi_{m}\pi\right\} = \sum_{x^{n} \in \CalX^{n}} p_{X}^{n}(x^{n})\left(\frac{q_{X}^{n}(x^{n})}{p_{X}^{n}(x^{n})}\right)^{2}\pi\pi_{x^{n}}\rho_{x^{n}}\pi_{x^{n}}\pi\pi\pi_{x^{n}}\rho_{x^{n}}\pi_{x^{n}}\pi \nonumber\\
 \label{Eqn:GenQCLSimplifyingT2_11}
&&\!\!\!\!\!\!\!\!\!\!\!\!\!\!\leq  \!\!\!\sum_{x^{n} \in T_{\eta}^{n}(q_{X})} \!\!\!\left(\frac{q_{X}^{n}(x^{n})}{\sqrt{p_{X}^{n}(x^{n})}}\right)^{2}\pi\pi_{x^{n}}\rho_{x^{n}}\pi_{x^{n}}\pi_{x^{n}}\rho_{x^{n}}\pi_{x^{n}}\pi~~\leq  \!\!\!\sum_{x^{n} \in T_{\eta}^{n}(q_{X})} \!\!\!\left(\frac{q_{X}^{n}(x^{n})}{\sqrt{p_{X}^{n}(x^{n})}}\right)^{2}\pi\pi_{x^{n}}\rho_{x^{n}}^{2}\pi_{x^{n}}\pi \\
  \label{Eqn:GenQCLSimplifyingT2_13}
 &&\!\!\!\!\!\!\!\!\!\!\!\!\!\!= \!\!\!\sum_{x^{n} \in T_{\eta}^{n}(q_{X})} \!\!\!\left(\frac{q_{X}^{n}(x^{n})}{\sqrt{p_{X}^{n}(x^{n})}}\right)^{2}\pi\rho_{x^{n}}^{\frac{1}{2}}\pi_{x^{n}}\rho_{x^{n}}\pi_{x^{n}}\rho_{x^{n}}^{\frac{1}{2}}\pi ~~\leq  \!\!\!\sum_{x^{n} \in T_{\eta}^{n}(q_{X})}\!\!\! 2^{-n\left(\sum_{x\in \CalX}q_{x}(x)S(\rho_{x})-\eta\right)}\left(\frac{q_{X}^{n}(x^{n})}{\sqrt{p_{X}^{n}(x^{n})}}\right)^{2}\pi\rho_{x^{n}}^{\frac{1}{2}}\pi_{x^{n}}\rho_{x^{n}}^{\frac{1}{2}}\pi\\
   \label{Eqn:GenQCLSimplifyingT2_15}
  \lefteqn{\!\!\!\!\leq \!\!\!\sum_{x^{n} \in T_{\eta}^{n}(q_{X})}\!\!\!\!\!\!\!\! 2^{-n\left(\sum_{x\in \CalX}q_{x}(x)S(\rho_{x})-\eta\right)}\!\!\left(\!\!\frac{q_{X}^{n}(x^{n})}{\sqrt{p_{X}^{n}(x^{n})}}\!\!\right)^{\!2}\!\!\!\pi\rho_{x^{n}}\pi~
  \leq \!\!\!\!\!\sum_{x^{n} \in T_{\eta}^{n}(q_{X})} \!\!\!\!\!\!2^{-n\left(\sum_{x\in \CalX}q_{x}(x)S(\rho_{x})-D(q_{X}||p_{X})[1 +4\eta]\right)}{q_{X}^{n}(x^{n})}\pi\rho_{x^{n}}\pi}\\
   \label{Eqn:GenQCLSimplifyingT2_18}
  &&\!\!\!\!\!\!\!\!\!\!\!\!\!\! \leq 2^{-n\left(\sum_{x\in \CalX}q_{x}(x)S(\rho_{x})-D(q_{X}||p_{X}) [1 +4\eta]\right)}\pi\sigma^{\otimes n}\pi~~\leq~~ 2^{-n\left(\sum_{x\in \CalX}q_{x}(x)S(\rho_{x})-D(q_{X}||p_{X})[1 +4\eta]+S(\rho) \right)}\pi
\end{eqnarray}
where (i) the restricted sum in the first inequality of \eqref{Eqn:GenQCLSimplifyingT2_11} follows from our definition of the conditional typical projector $\pi_{x^{n}}$ being the zero operator if $x^{n} \notin T_{\eta}^{n}(q_{X})$, (ii) the equality in \eqref{Eqn:GenQCLSimplifyingT2_13} follows from the fact that $\pi_{x^{n}}$ and $\rho_{x^{n}}$ commute, (iii) the inequality in \eqref{Eqn:GenQCLSimplifyingT2_13} follows from the operator bound $\pi_{x^{n}}\rho_{x^{n}}\pi_{x^{n}} \leq 2^{-n\left(\sum_{x\in \CalX}q_{x}(x)S(\rho_{x})-\eta\right)}\pi_{x^{n}}$ for $x^{n} \in T_{\eta}^{n}(q_{X})$ found in \cite[Property 15.2.6]{BkWilde_2017}, (iv) the first inequality in \eqref{Eqn:GenQCLSimplifyingT2_15} follows again from $A\pi_{x^{n}} A^{\dagger} \leq AA^{\dagger}$ for any $A$, (v) the second inequality in \eqref{Eqn:GenQCLSimplifyingT2_15} follows from the standard typicality bound which states that for $x^{n} \in T_{\eta}^{n}(x^{n})$ we have $\frac{q_{X}^{n}(x^{n})}{p_{X}^{n}(x^{n})} \leq \exp\{ -n [H(q_{X}) -D(q_{X}||p_{X})(1+2\eta)-H(q_{X})-4\eta ] \}$, (vi) the first inequality in \eqref{Eqn:GenQCLSimplifyingT2_18} follows from enlarging the sum of non-negative operators $q_{X}^{n}(x^{n})\pi\rho_{x^{n}}\pi$ from $T_{\eta}^{n}(q_{X})$ to $\CalX^{n}$ and recognizing that $\sum_{x^{n}\in \CalX^{n}}q_{X}(x^{n})\rho_{x^{n}} = \sigma^{\otimes n}$, and (vii) follows from the fact that $\pi$ is the unconditional typical projector of $\sigma$ and the operator bound $\pi\sigma^{\otimes n}\pi \leq 2^{-nS(\sigma)+2n\eta}$ found in \cite[Property 15.1.3]{BkWilde_2017}. Substituting the upper bound in \eqref{Eqn:GenQCLSimplifyingT2_18} for each term in the sum under the square root in \eqref{Eqn:GenQCLSimplifyingT2_9} and recognizing that the upper bound is invariant with $m \in [M]$, we have 
\begin{eqnarray}
 \label{Eqn:GenQCLSimplifyingT2_19}
 T_{2} \leq  \exp\{ -\frac{n}{2}\left( \frac{\log M}{n} +\sum_{x\in \CalX}q_{x}(x)S(\rho_{x})-D(q_{X}||p_{X})[1 +4\eta]+S(\rho)  \right) \}\tr(\sqrt{\pi}) \nonumber\\
 \label{Eqn:GenQCLSimplifyingT2_20}
 \leq \exp\left\{ -\frac{n}{2}\left( \frac{\log M}{n} +\sum_{x\in \CalX}q_{x}(x)S(\rho_{x})-D(q_{X}||p_{X})[1 +4\eta]-S(\rho)  \right) \right\}
\end{eqnarray}
This completes our analysis of $T_{2}$ and we are left with $T_{1},T_{3}$. Before we analyze the latter terms separately, we set up some notation and recall some facts. For the random collection $C=(X^{n}(m) \in T_{\eta}^{n}(q_{X}):1\leq m \leq M)$, we let
\begin{eqnarray}
 \label{Eqn:QCLNotationsForAAndB1}
 A_{t}(m) \define A(m)\mathds{1}_{\{ X^{n}(m) \in T_{{\eta}}^{n}(q_{X})\}},~ A_{nt}(m) \define A(m)\mathds{1}_{\{ X^{n}(m) \notin T_{{\eta}}^{n}(q_{X})\}},~B_{t}(m) \define B(m)\mathds{1}_{\{  X^{n}(m) \in T_{{\eta}}^{n}(q_{X}) \}}, \nonumber\\
 \label{Eqn:QCLNotationsForAAndB2}
 B_{nt}(m) \define B(m)\mathds{1}_{\{  X^{n}(m) \notin T_{{\eta}}^{n}(q_{X}) \}} \mbox{ and note } A(m)= A_{t}(m)+A_{nt}(m), B(m)= B_{t}(m)+B_{nt}(m).~~~~
 \nonumber
\end{eqnarray}
We now make note of a fact that we will have opportunity to use on multiple occasions. From the gentle measurement lemma \cite[Lemma~9.4.2]{BkWilde_2017}, we have $\norm{\rho_{x^{n}} - \pi_{x^{n} } \rho_{x^{n}}\pi_{x^{n}}}_{1} \leq 2\sqrt{\tr([I-\pi_{x^{n}}]\rho_{x^{n}})}$. Next, from the measurement on close states \cite[Corollary 9.1.1, Exercise 9.1.8]{BkWilde_2017} and combining the above inequality, we have
\begin{eqnarray}
\label{Eqn:RepeatedGentleMsmt1}
 \tr(\pi\pi_{x^{n}}\rho_{x^{n}}\pi_{x^{n}}) \geq \tr(\pi\rho_{x^{n}})  - \norm{\rho_{x^{n}} - \pi_{x^{n} } \rho_{x^{n}}\pi_{x^{n}}}_{1} \geq \tr(\pi\rho_{x^{n}})  - 2\sqrt{\tr([I-\pi_{x^{n}}]\rho_{x^{n}})}.
 \end{eqnarray}
Once again, leveraging the gentle measurement lemma \cite[Lemma~9.4.2]{BkWilde_2017}, this time on the `sub-normalized state' $\pi_{x^{n}}\rho_{x^{n}}\pi_{x^{n}}$ \cite[Exercise 9.4.1]{BkWilde_2017}, we have
\begin{eqnarray}
 \norm{\pi_{x^{n}}\rho_{x^{n}}\pi_{x^{n}} - \pi\pi_{x^{n}}\rho_{x^{n}}\pi_{x^{n}}\pi}_{1} &\leq& 2\sqrt{\tr([I-\pi]\pi_{x^{n}}\rho_{x^{n}}\pi_{x^{n}} )} = 2\sqrt{\tr(\rho_{x^{n}}\pi_{x^{n}}) - \tr(\pi\pi_{x^{n}}\rho_{x^{n}}\pi_{x^{n}})} \nonumber\\ 
 \label{Eqn:RepeatedGentleMsmt2}
 &\leq& 4\sqrt{\tr(\pi_{x^{n}}\rho_{x^{n}})-\tr(\pi\rho_{x^{n}})  + 2\sqrt{\tr([I-\pi_{x^{n}}]\rho_{x^{n}})}},
\end{eqnarray}
where we utilized \eqref{Eqn:RepeatedGentleMsmt1} in the last inequality. Combining \eqref{Eqn:RepeatedGentleMsmt2} with the bound on $\norm{\rho_{x^{n}} - \pi_{x^{n} } \rho_{x^{n}}\pi_{x^{n}}}_{1}$ derived prior to \eqref{Eqn:RepeatedGentleMsmt1} and using the triangular inequality, we have
\begin{eqnarray}
 \label{Eqn:RepeatedGentleMsmtCombined}
 \norm{\rho_{x^{n}} - \pi\pi_{x^{n}}\rho_{x^{n}}\pi_{x^{n}}\pi}_{1} \leq 2\sqrt{\tr([I-\pi_{x^{n}}]\rho_{x^{n}})}+ 4\sqrt{\tr(\pi_{x^{n}}\rho_{x^{n}})-\tr(\pi\rho_{x^{n}})  + 2\sqrt{\tr([I-\pi_{x^{n}}]\rho_{x^{n}})}}.
\end{eqnarray}
For $x^{n} \in T_{\eta}(q_{X})$ we conclude from the properties of conditional and unconditional typical projectors \cite[Properties 15.2.4 and 15.2.7]{BkWilde_2017} that
\begin{eqnarray}
 \label{Eqn:RepeatedGentleMsmt}
 \norm{\rho_{x^{n}} - \pi\pi_{x^{n}}\rho_{x^{n}}\pi_{x^{n}}\pi}_{1} \leq 2^{-n\frac{\eta}{10}}
\end{eqnarray}
for sufficiently large $n$. We are now set to analyze $T_{1}$.
\med\textit{An Upper Bound on $T_{1}$ : }Since $A=\frac{1}{M}\sum_{m=1}^{M}A(m) = \frac{1}{M}\sum_{m=1}^{M}[A_{t}(m) +A_{nt}(m)] $ and $B=\frac{1}{M}\sum_{m=1}^{M}B(m) = \frac{1}{M}\sum_{m=1}^{M}[B_{t}(m) +B_{nt}(m)] $, we have
\begin{eqnarray}
  T_{1} \!\!&\!\!\!=\!\!\!&\!\!  \Expectation_{\mathbf{P}}\left\{\norm{ A-B}_{1} \right\} \leq 
\label{Eqn:GenQCLSimplifyingT1_2}
 \frac{1}{M}\sum_{m=1}^{M}\Expectation_{\mathbf{P}}\left\{ \norm{A_{t}(m) - B_{t}(m)}_{1}  \right\}+\frac{1}{M}\sum_{m=1}^{M}\Expectation_{\mathbf{P}}\left\{ \norm{A_{nt}(m) - B_{nt}(m)}_{1}  \right\}\\
 \label{Eqn:GenQCLSimplifyingT1_3}
\!\!&\!\!\!= \!\!\!&\!\! \frac{1}{M}\sum_{m=1}^{M}\Expectation_{\mathbf{P}}\left\{ \frac{q_{X}^{n}(X^{n}(m))}{p_{X}^{n}(X^{n}(m))}\norm{\rho_{X^{n}(m)}\mathds{1}_{\{ X^{n}(m) \in T_{{\eta}}^{n}(q_{X})\}}  - \pi\pi_{m}\rho_{X^{n}(m)}\pi_{m}\pi\mathds{1}_{\{ X^{n}(m) \in T_{{\eta}}^{n}(q_{X})\}} }_{1}  \right\}\nonumber\\
 \label{Eqn:GenQCLSimplifyingT1_4}
&&~~~~~~+\frac{1}{M}\sum_{m=1}^{M}\Expectation_{\mathbf{P}}\left\{\frac{q_{X}^{n}(X^{n}(m))}{p_{X}^{n}(X^{n}(m))} \norm{\rho_{X^{n}(m)}\mathds{1}_{\{ X^{n}(m) \notin T_{{\eta}}^{n}(q_{X})\}}  - \pi\pi_{m}\rho_{X^{n}(m)}\pi_{m}\pi\mathds{1}_{\{ X^{n}(m) \notin T_{{\eta}}^{n}(q_{X})\}}}_{1}  \right\}\\
\label{Eqn:GenQCLSimplifyingT1_5}
\!\!&\!\!\!\leq \!\!\!&\!\! \frac{1}{M}\sum_{m=1}^{M}\Expectation_{\mathbf{P}}\left\{ \frac{q_{X}^{n}(X^{n}(m))}{p_{X}^{n}(X^{n}(m))}2^{-n\frac{\eta}{10}}  \right\}+ \frac{1}{M}\sum_{m=1}^{M}\Expectation_{\mathbf{P}}\left\{\frac{q_{X}^{n}(X^{n}(m))}{p_{X}^{n}(X^{n}(m))} \mathds{1}_{\{ X^{n}(m) \notin T_{{\eta}}^{n}(q_{X})\}}\norm{\rho_{X^{n}(m)}}_{1}\right\}  \nonumber\\ 
 \label{Eqn:GenQCLSimplifyingT1_6}
&&~~~~~+\frac{1}{M}\sum_{m=1}^{M}\Expectation\left\{\frac{q_{X}^{n}(X^{n}(m))}{p_{X}^{n}(X^{n}(m))}\mathds{1}_{\{ X^{n}(m) \notin T_{{\eta}}^{n}(q_{X})\}}\norm{ \pi\pi_{m}\rho_{X^{n}(m)}\pi_{m}\pi}_{1}  \right\}\\
\label{Eqn:GenQCLSimplifyingT1_7}
\!\!&\!\!\!\leq \!\!\!&\!\! \frac{1}{M}\sum_{m=1}^{M}\sum_{u^{n} \in \CalX^{n}} \frac{q_{X}^{n}(u^{n})}{p_{X}^{n}(u^{n})}p_{X}^{n}(u^{n})2^{-n\frac{\eta}{10}}  + \frac{1}{M}\sum_{m=1}^{M}\Expectation_{\mathbf{P}}\left\{\frac{q_{X}^{n}(X^{n}(m))}{p_{X}^{n}(X^{n}(m))} \mathds{1}_{\{ X^{n}(m) \notin T_{{\eta}}^{n}(q_{X})\}}1\right\}\nonumber\\ 
 \label{Eqn:GenQCLSimplifyingT1_10}
&&~~~~~+\frac{1}{M}\sum_{m=1}^{M}\Expectation\left\{\frac{q_{X}^{n}(X^{n}(m))}{p_{X}^{n}(X^{n}(m))}\mathds{1}_{\{ X^{n}(m) \notin T_{{\eta}}^{n}(q_{X})\}}\norm{ \pi}_{\infty}\norm{\pi_{m}}_{\infty}\norm{\rho_{X^{n}(m)}}_{1}\norm{\pi_{m}}_{\infty}\norm{\pi}_{\infty}  \right\}\\
\label{Eqn:GenQCLSimplifyingT1_11}
\!\!&\!\!\!\leq \!\!\!&\!\! 2^{-n\frac{\eta}{10}}  + \frac{2}{M}\sum_{m=1}^{M} \sum_{u^{n}}\frac{q_{X}^{n}(u^{n})}{p_{X}^{n}(u^{n})}p_{X}(u^{n}) \mathds{1}_{\{ u^{n} \notin T_{{\eta}}^{n}(q_{X})\}} \leq2^{-n\frac{\eta}{10}}+2\cdot 2^{-n\frac{\eta}{20}}
\end{eqnarray}
for sufficiently large $n$, where (i) \eqref{Eqn:GenQCLSimplifyingT1_2} follows from the triangular inequality, (ii) \eqref{Eqn:GenQCLSimplifyingT1_4} follows from substitution and pulling the common non-negative scalar factors out of the norm, (iii) \eqref{Eqn:GenQCLSimplifyingT1_6} follows from \eqref{Eqn:RepeatedGentleMsmt}, the triangular inequality and pulling the non-negative indicator functions out of the norm, (iv) \eqref{Eqn:GenQCLSimplifyingT1_10} follows from $\norm{\rho_{x^{n}}}_{1}=\tr(\rho_{x^{n}})=1$ and the repeated use of the bound $\norm{AB}_{1} \leq \norm{A}_{\infty}\norm{B}_{1}$ and (v) follows from the fact that $\norm{ \pi}_{\infty}=\norm{\pi_{m}}_{\infty}=1 $, $\norm{\rho_{x^{n}}}_{1}=\tr(\rho_{x^{n}})=1$ implying that the second and third terms of \eqref{Eqn:GenQCLSimplifyingT1_10} are identical, which is then bounded above via typicality arguments.
\med\textit{Upper Bound on $T_{3}$} : We note that
\begin{eqnarray}
\label{Eqn:GenQCLSimplifyingT3_1}
 \norm{\overline{A}-\overline{B}}_{1} \!\!&\!\!=\!\!&\!\! \norm{\frac{1}{M}\sum_{m=1}^{M}\Expectation_{\mathbf{P}}\left\{ A_{t}(m)-B_{t}(m)\right\}+\frac{1}{M}\sum_{m=1}^{M}\Expectation_{\mathbf{P}}\left\{ A_{nt}(m)-B_{nt}(m)\right\}}_{1} \nonumber\\
 \label{Eqn:GenQCLSimplifyingT3_2}
 \!\!&\!\!\leq\!\!&\!\!\frac{1}{M}\sum_{m=1}^{M}\norm{\Expectation_{\mathbf{P}}\left\{ A_{t}(m)-B_{t}(m)\right\}}_{1}+\frac{1}{M}\sum_{m=1}^{M}\norm{\Expectation_{\mathbf{P}}\left\{ A_{nt}(m)-B_{nt}(m)\right\}}_{1}
 \\
 \label{Eqn:GenQCLSimplifyingT3_3}
 \!\!&\!\!=\!\!&\!\!\frac{1}{M}\sum_{m=1}^{M}\norm{\Expectation_{\mathbf{P}}\left\{ \frac{q_{X}^{n}(X^{n}(m))}{p_{X}^{n}(X^{n}(m))}\left[\rho_{X^{n}(m)}\mathds{1}_{\{ X^{n}(m) \in T_{{\eta}}^{n}(q_{X})\}}  - \pi\pi_{m}\rho_{X^{n}(m)}\pi_{m}\pi\mathds{1}_{\{ X^{n}(m) \in T_{{\eta}}^{n}(q_{X})\}}\right]   \right\}}_{1}\nonumber\\
 \label{Eqn:GenQCLSimplifyingT3_4}
 &&+\frac{1}{M}\sum_{m=1}^{M}\norm{\Expectation_{\mathbf{P}}\left\{ \frac{q_{X}^{n}(X^{n}(m))}{p_{X}^{n}(X^{n}(m))}\left[\rho_{X^{n}(m)}\mathds{1}_{\{ X^{n}(m) \notin T_{{\eta}}^{n}(q_{X})\}}  - \pi\pi_{m}\rho_{X^{n}(m)}\pi_{m}\pi\mathds{1}_{\{ X^{n}(m) \notin T_{{\eta}}^{n}(q_{X})\}}\right]   \right\}}_{1}\nonumber\\
 \label{Eqn:GenQCLSimplifyingT3_5}
 \!\!&\!\!=\!\!&\!\!\frac{1}{M}\sum_{m=1}^{M}\norm{\sum_{u^{n} \in \CalX^{n}} \frac{q_{X}^{n}(u^{n})}{p_{X}^{n}(u^{n})}p_{X}(u^{n})\left[\rho_{u^{n}}\mathds{1}_{\{ u^{n} \in T_{{\eta}}^{n}(q_{X})\}}  - \pi\pi_{m}\rho_{u^{n}}\pi_{m}\pi\mathds{1}_{\{ u^{n} \in T_{{\eta}}^{n}(q_{X})\}}\right]  }_{1}\nonumber\\
 \label{Eqn:GenQCLSimplifyingT3_6}
 &&+\frac{1}{M}\sum_{m=1}^{M}\norm{\sum_{v^{n} \in \CalX^{n}} \frac{q_{X}^{n}(v^{n})}{p_{X}^{n}(v^{n})}p_{X}^{n}(v^{n})\left[\rho_{v^{n}}\mathds{1}_{\{ v^{n} \notin T_{{\eta}}^{n}(q_{X})\}}  - \pi\pi_{m}\rho_{v^{n}}\pi_{m}\pi\mathds{1}_{\{ v^{n} \notin T_{{\eta}}^{n}(q_{X})\}}\right]   }_{1}\nonumber\\
 \label{Eqn:GenQCLSimplifyingT3_7}
 \!\!&\!\!\leq\!\!&\!\!\frac{1}{M}\sum_{m=1}^{M}\sum_{u^{n} \in \CalX^{n}} q_{X}(u^{n})\norm{\left[\rho_{u^{n}}\mathds{1}_{\{ u^{n} \in T_{{\eta}}^{n}(q_{X})\}}  - \pi\pi_{m}\rho_{u^{n}}\pi_{m}\pi\mathds{1}_{\{ u^{n} \in T_{{\eta}}^{n}(q_{X})\}}\right]  }_{1}\nonumber\\
 \label{Eqn:GenQCLSimplifyingT3_8}
 &&+\frac{1}{M}\sum_{m=1}^{M}\sum_{v^{n} \in \CalX^{n}} q_{X}^{n}(v^{n})\norm{\left[\rho_{v^{n}}\mathds{1}_{\{ v^{n} \notin T_{{\eta}}^{n}(q_{X})\}}  - \pi\pi_{m}\rho_{v^{n}}\pi_{m}\pi\mathds{1}_{\{ v^{n} \notin T_{{\eta}}^{n}(q_{X})\}}\right]   }_{1}\\
 \label{Eqn:GenQCLSimplifyingT3_9}
\!\!&\!\!\leq\!\!&\!\!\frac{1}{M}\sum_{m=1}^{M}\sum_{u^{n} \in \CalX^{n}} q_{X}(u^{n})2^{-n\frac{\eta}{10}}++\frac{2}{M}\sum_{m=1}^{M}\sum_{v^{n} \in \CalX^{n}} q_{X}^{n}(v^{n})\mathds{1}_{\{ v^{n} \notin T_{{\eta}}^{n}(q_{X})\}} \leq 2^{-n\frac{\eta}{5}}
\end{eqnarray}
for sufficiently large $n$, where (i) \eqref{Eqn:GenQCLSimplifyingT3_2}, \eqref{Eqn:GenQCLSimplifyingT3_8} follow from triangular inequality and (ii) \eqref{Eqn:GenQCLSimplifyingT3_9} follows from \eqref{Eqn:RepeatedGentleMsmt} and 
\begin{eqnarray}
\lefteqn{\norm{\left[\rho_{v^{n}}\mathds{1}_{\{ v^{n} \notin T_{{\eta}}^{n}(q_{X})\}}  - \pi\pi_{m}\rho_{v^{n}}\pi_{m}\pi\mathds{1}_{\{ v^{n} \notin T_{{\eta}}^{n}(q_{X})\}}\right]   }_{1} = \mathds{1}_{\{ v^{n} \notin T_{{\eta}}^{n}(q_{X})\}}\norm{\left[\rho_{v^{n}}  - \pi\pi_{m}\rho_{v^{n}}\pi_{m}\pi\right]   }_{1}} \nonumber\\ &&\leq \mathds{1}_{\{ v^{n} \notin T_{{\eta}}^{n}(q_{X})\}}\left(\norm{\rho_{v^{n}}}_{1}  + \norm{\pi\pi_{m}\rho_{v^{n}}\pi_{m}\pi   }_{1}\right) \leq \mathds{1}_{\{ v^{n} \notin T_{{\eta}}^{n}(q_{X})\}}\left(1 + \norm{\pi\pi_{m}\rho_{v^{n}}\pi_{m}\pi   }_{1}\right)  \leq 2\mathds{1}_{\{ v^{n} \notin T_{{\eta}}^{n}(q_{X})\}}.
\nonumber
\end{eqnarray}
Here, the last inequality above follows from the same argument we provided in going from \eqref{Eqn:GenQCLSimplifyingT1_6} to \eqref{Eqn:GenQCLSimplifyingT1_11}.

The proof of the theorem follows by combining the bounds derived in  \eqref{Eqn:GenQCLSimplifyingT1_11} and \eqref{Eqn:GenQCLSimplifyingT3_9}.
\end{proof}

\bibliographystyle{IEEEtran}
{
\bibliography{Quant_InfExtNtwrkPOVMs}

\end{document}